\documentclass[11pt]{article}
\usepackage[utf8]{inputenc}
\usepackage{authblk}
\usepackage{graphicx}
\usepackage{xcolor}
\usepackage[round]{natbib}
\usepackage{aas_macros}
\usepackage{lineno,hyperref}

\def\sun{\hbox{$_\odot$}}

\def\ga{\mathrel{\mathchoice {\vcenter{\offinterlineskip\halign{\hfil
$\displaystyle##$\hfil\cr>\cr\sim\cr}}}
{\vcenter{\offinterlineskip\halign{\hfil$\textstyle##$\hfil\cr
>\cr\sim\cr}}}
{\vcenter{\offinterlineskip\halign{\hfil$\scriptstyle##$\hfil\cr
>\cr\sim\cr}}}
{\vcenter{\offinterlineskip\halign{\hfil$\scriptscriptstyle##$\hfil\cr
>\cr\sim\cr}}}}}

\def\utw{\smash{\rlap{\lower5pt\hbox{$\sim$}}}}
\def\udtw{\smash{\rlap{\lower6pt\hbox{$\approx$}}}}

\def\farcs{\hbox{${.}\!\!^{\prime\prime}$}}

\newcommand{\Hii}{H~{\sc ii}}
\newcommand{\Hi}{H~{\sc i}}
\newcommand{\kms}{km\,s$^{-1}$}
\newcommand{\pcmc}{cm$^{-3}$}
\newcommand{\pcms}{cm$^{-2}$}

\title{Observational studies of high-mass star formation}
\author{Igor I. Zinchenko}
\affil{Institute of Applied Physics of the Russian Academy of Science\\
46 Ul'yanov str., Nizhny Novgorod, Russia\\
E-mail: \href{mailto:zin@iapras.ru}{zin@iapras.ru}}

\begin{document}

\maketitle

\begin{abstract}
\noindent
We present a review of observational studies of high-mass star formation, based mainly on our own research. It includes surveys of high-mass star-forming regions in various molecular lines and in continuum, investigations of filamentary infrared dark clouds, which represent the earliest phases of massive star formation, detailed studies of individual high-mass star-forming regions, dense cores and disks harboring massive (proto)stars, and associated outflows. Chemistry in these regions is discussed, too.


\end{abstract}

\section{Introduction}
Massive stars ($ M\ga 8 $~M\sun) play a very important role in the life of the Universe. However the process of their formation is still puzzling in many respects. Both theoretical and observational studies of this process face serious difficulties. The process of formation of massive stars seems to be more complicated than the formation of low-mass stars, and there are still many unclear points in it \citep[e.g.][]{McKee07, Zinnecker07, Tan14, Motte18, Rosen20} related to the fact that nuclear reactions in massive protostars begin much earlier than they reach the final mass. Radiation pressure can stop further influx of matter. It is also not entirely clear how to explain the fact that massive cores do not break up into smaller fragments. Observational studies of HMSF regions are hampered by the fact that they are rare and are located much farther from us than the dark clouds in which stars of small mass are formed. The nearest such region is at a distance of $ \sim 500$~pc, and typical distances are several kiloparsecs. Interferometers are required for their detailed study. 

Here we present a brief review of observational studies of high-mass star formation with an emphasis on our own results.

\section{Surveys of star-forming regions in the Milky Way galaxy} \label{sec:surveys}
Studies of the general characteristics of star-forming regions are based on surveys of these objects. To date, quite a few such surveys have been carried out.
A good example of such work is the survey of the Galactic plane in the CO $ J=1-0 $ line \citep{Dame01}, which served as the basis for many further studies. Such a survey gives a general idea of the distribution and kinematics of interstellar matter, but has a rather low angular resolution and makes it difficult to identify and study compact star formation regions. For such problems, surveys of significant parts of the Galactic plane with a much higher resolution in the continuum and in the lines of some molecules, carried out with the help of ground-based and space instruments, are very useful. Among them, one can note ground surveys in the continuum at wavelengths $\sim 1$~mm ATLASGAL (The APEX telescope large area survey of the galaxy at 870 {$\mu$}m) \citep{Schuller09,Csengeri14} and BOLOCAM \citep{Aguirre11}, as well as the survey in the $^{13}$CO $ J=1-0 $ line GRS (The Boston University-Five College Radio Astronomy Observatory Galactic Ring Survey) \citep{Jackson06}. A lot of useful information was given by spacecraft operating at wavelengths from the far IR to the near IR ranges (for example, \textit{Spitzer} Galactic Legacy Infrared Mid-Plane Survey Extraordinaire --- GLIMPSE \citealt{Benjamin03} and MIPS Inner Galactic Plane Survey --- MIPSGAL \citealt{Carey05}, \textit{Herschel} \citealt{Pilbratt10} infrared Galactic Plane Survey --- Hi-GAL \citealt{Molinari16}, Wide Field Infrared Survey Explorer --- WISE \citealt{Wright10}). The results of these surveys are now actively used in the study of star-forming regions.

Another type of such work is surveys of samples of objects selected according to certain criteria, which may represent various types of star-forming regions or may be indicators of such regions. Surveys are carried out in the continuum at millimeter and submillimeter wavelengths, as well as in the lines of common molecules such as CO, CS, NH$_3$, HCN, HCO$^+$, N$_2$H$^+$ and some others. As a prominent example of such studies, one can note a series of works by Philip Myers and co-authors on the study of dense cores in dark clouds, carried out quite a long time ago \citep{DCDCL1, DCDCL2, DCDCL3, DCDCL4, DCDCL5, DCDCL6, DCDCL7, DCDCL8, DCDCL9, DCDCL10, DCDCL11, DCDCL12, DCDCL14, Myers85, Benson89, Goodman90}. As a result of these works, the main physical characteristics of such cores were determined, the chemical composition was studied, and the correlations between the parameters were studied.

\begin{figure}
    \centering
    \includegraphics[width=0.9\textwidth]{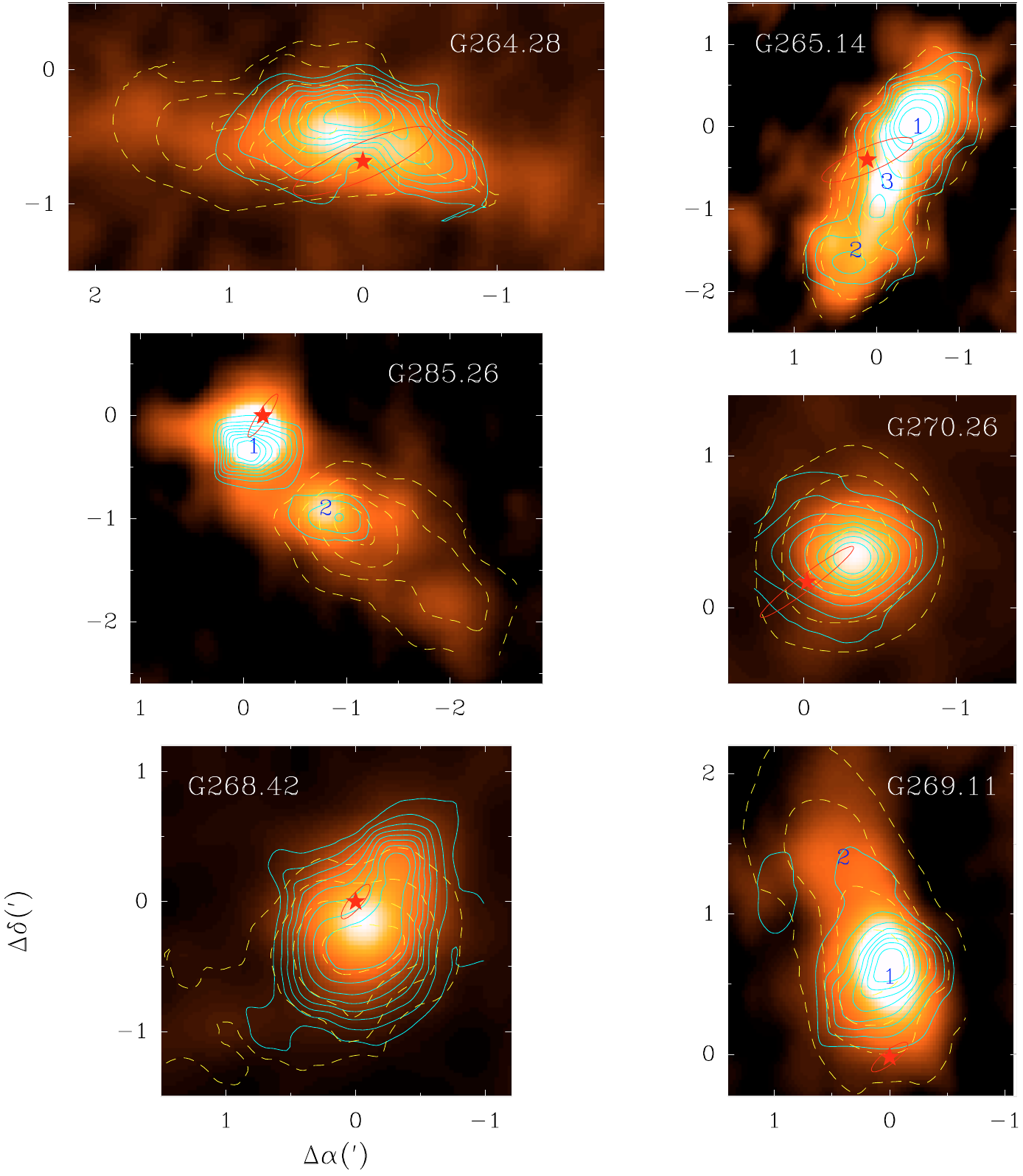}
    \caption{Continuum 1.2 mm maps (in color) together with CS(5--4) (solid blue contours) and N$_2$H$^+$(1--0) (dashed yellow contours, taken from \citealt{Pirogov03}) integrated intensity maps. The IRAS point sources are marked by red stars. The uncertainty ellipses corresponding to 95\% confidence level in the IRAS point-source position are also shown. The SEST beam size at the CS(5--4) frequency is shown on the G264.28 map. Adopted from \citet{Pirogov07}.}
    \label{fig:cs_n2h_survey}
\end{figure}

In these dark clouds, stars of low mass form, of the order of the mass of the Sun and less. Similar work has been carried out and is being carried out in the direction of HMSF regions. 
Among the many works on surveys of such areas, we note a series of our studies that began in the 80s of the last century on the RT-22 of the Crimean Astrophysical Observatory \citep{Burov88, Zin89, Zin90} and then continued on various instruments worldwide \citep{Zin94, Zin95, Zin95-2, Pirogov96, Zin97, Lapinov98, Zin98, Zin00, Pirogov03, Pirogov07, Zin09, Zin-SO}. In these papers, surveys were made of several dozen HMSF regions in the lines of such molecules as HCN, HCO$^+$, CS, NH$_3$, N$_2$H$^+$, HNCO, C$^{18 }$O, SO, etc. An example of such a survey is presented in Fig.~\ref{fig:cs_n2h_survey}. This made it possible to obtain statistical distributions of the main physical parameters for these objects, in particular, sizes, masses, density, velocity dispersion. The kinetic temperature of the gas  was estimated from the observations of NH$_3$ and CH$_3$CCH \citep{Zin97, Malafeev05, Malafeev06}.

One of the interesting results was the discovery of a decrease in the average cloud density with increasing galactocentric distance \citep{Zin95-2,Zin98}. Similar changes in the properties of the interstellar gas were also noted in other papers. For example, \citet{Sakamoto97} found a strong decrease in the ratio of CO line intensities $ J=2-1 $ and $ J=1-0 $ with distance from the center of the Galaxy, which most likely indicates a decrease in gas density (and possibly temperature). Such studies are best, of course, carried out on the basis of observations of neighboring galaxies. Observations of HCN and HCO$^+$ in M31 \citep{Brouillet05} demonstrate a noticeable drop in the intensity ratios $ I(\mathrm{HCN})/I(\mathrm{CO}) $ and $ I(\mathrm{HCO^+} )/I(\mathrm{CO}) $ with increasing galactocentric distance, which is interpreted as a result of changing the excitation conditions of molecules. With the advent of new tools, such work can give new important results.

\section{The earliest phases of massive star formation}
Various criteria can be used to identify the early stages of massive star formation. Shortly after the appearance of IRAS  survey data, \citet{Wood89} proposed a method for identifying massive stars embedded in molecular clouds (i.e., at a relatively early stage of evolution) by their emission spectrum in the far IR range (based on the so-called ``two-color diagrams''. A characteristic spectrum is given by the ultracompact \Hii\ zones associated with these stars.

Further it was shown that earlier stages can be marked by maser emission of some molecules (H$_2$O, CH$_3$OH) \citep{Plume97,Walsh97}. A number of surveys of maser sources in the lines of various molecules, including ours \citep{Zin95,Zin98}, have been carried out, which made it possible to determine the main characteristics of molecular clumps associated with these masers.

A number of authors \citep[e.g.][]{Molinari96,Sridharan02} proposed to use approximately the same IR colors criterion, but with the additional condition of the absence of radio emission in the continuum at relatively low frequencies (which is usually generated due to bremsstrahlung of ionized gas). Thus, one can hope to identify objects that are at the stage preceding the formation of the ultracompact \Hii\ zone.

Samples of objects that meet these criteria have been actively studied, but it is obvious that in all of them the process of star formation is already underway. And one would like to find for massive stars some kind of pre-stellar cloud, where it has not yet begun. For low-mass stars, such prestellar objects are dark cold clouds, which are located relatively close to us ($\sim$100~pc) and are observed in optics as dark dips against the background of stars. More than 20 years ago, infrared dark clouds were discovered, which are probably the objects where stars of large mass will be formed in the future or are already beginning to form.

In much of the infrared range, observations from the Earth's surface are practically impossible. Infrared astronomy began to develop actively only with the advent of appropriate space facilities. In the mid-1990s, images of a significant part of the Galactic plane were obtained in the mid-IR range. These images revealed a large amount of dark details \citep{Perault96,Egan98}, which, obviously, were created by fairly dense cold clouds that absorb background radiation from the disk of the Galaxy. It was immediately suggested that they are the sought-after massive prestellar objects in which stars of high mass can form. In recent years, infrared dark clouds (IRDC) have been the subject of detailed studies in various wavelength ranges, thanks to which their main physical characteristics and chemical composition have been determined.

The first molecular observations of infrared dark clouds (in formaldehyde lines \citep{Carey98}) showed that they are far from us -- at distances from 1 to 8\,kpc, and their sizes range from 0.4 to 15\,pc. Formaldehyde excitation analysis gave density ($ n>10^5$~cm$^{-3}$) and temperature ($ T<20$~K) estimates.

Further studies in the continuum and in molecular lines made it possible to refine the values of the physical parameters of the clouds. It was found, in particular, that their masses are from hundreds to thousands of solar masses, the H$_2$ column density is from 2 to $10\times 10^{23}$~cm$^{-2}$ \citep[e.g.][]{Rathborne06,Vasyunina09,Ragan09}.

Reliable temperature estimates (10--20~K) have been obtained from observations of lines of ammonia \citep{Pillai06}, which, like a number of other molecules of the symmetrical top type, is a good indicator of temperature in dense interstellar clouds. The linewidths, which are determined by the dispersion of gas velocities, range from $\sim$0.5 to $\sim$3~\kms\ in IRDC \citep[e.g.][]{Vasyunina11}. In terms of this parameter, infrared dark clouds occupy an intermediate position between cold dark clouds of low mass and warm massive clouds that already contain high-luminosity young stars.

Much attention is paid to the study of the chemical composition of infrared dark clouds. Features of this composition may indicate their evolutionary status, since this composition must change over time. The results of these studies are somewhat contradictory so far. Based on very low values of the abundance ratio CCS/N$_2$H$^+$, \citet{Sakai08} concluded that infrared dark clouds are at a later stage of chemical evolution than prestellar low-mass clouds. At the same time, \citet{Gibson09}, from the analysis of data on the abundance of C$^{18}$O, CS and N$_2$H$^+$, found that the chemical age of clouds is small and in some cases does not exceed 100 years. Studies of 15 infrared dark clouds in the lines of 13 different molecules \citet{Vasyunina11} have shown that they are closer in chemical composition to low-mass prestellar clouds than to massive protostellar objects (clouds containing young massive stars).


Many Infrared Dark Clouds are filamentous \citep{Schisano20}. 
Filamentous structures have long been observed and studied in the interstellar medium \citep[see, for example, reviews in][]{Andre14, Myers09}. They are visible in optical images of dark nebulae, in the structure of clouds of neutral atomic hydrogen observed in the 21~cm line, in molecular clouds of various types. 

In recent years, thanks to the emergence of a large amount of new observational data, primarily obtained from the \textit{Herschel} space observatory \citep{Pilbratt10}, it has become clear that interstellar filaments play a key role in the formation of stars \citep{Andre14}. The formation of filaments apparently precedes the onset of active star formation, since a filamentous structure is also observed in clouds where there are no signs of this process \citep[for example, Polaris Flare --][]{Ward-Thompson10}.

Studies of filaments at the \textit{Herschel} observatory were limited to objects relatively close to us. Subsequently, investigations of filamentous structures on the scales of the Galaxy were carried out, in particular, based on the analysis of data from the ASTROGAL survey \citep{Li16}. In this work, about 500 filamentous structures were identified at distances up to $ \sim 12 $~kpc. To select filaments on extended maps of interstellar clouds, automated algorithms such as \textit{getfilaments} \citep{Menshchikov13} are usually used. 

The observed filaments exhibit a wide range of physical parameters. At the same time, a characteristic feature of filaments in clouds close to us is the constancy of their width ($ \sim 0.1 $~pc) in different objects \citep{Andre14}. However, the more distant filaments identified in the \citet{Li16} survey mentioned above have, on average, a substantially larger width $ \sim 0.5 $~pc. At the same time, much narrower structures with a width of $ \sim 0.035 $~pc have also been identified in the Orion Nebula \citep{Hacar18}. In addition, some filaments have been found to be composed of many smaller fibers whose properties vary considerably \citep{Hacar13}. Some of these small filaments are actively star-forming, while others are not.

Almost perpendicular to the main filaments, thinner similar formations often adjoin, which are called striations. Polarization measurements show that the magnetic field, as a rule, is oriented perpendicular to the main filaments and, accordingly,
parallel to the striations \citep{Andre14}. It is assumed that along these striations there is an influx of substance onto the filaments. There are even smaller similar formations that can be called ``strands'' \citep{Cox16}. For the Musca filament studied in detail, the column density of hydrogen is $ N(\mathrm{H_2}) > 2.7 \times 10^{21} $~\pcms\ for the main filament, $ N(\mathrm{H_2}) < 1.8 \times 10 ^{21} $~\pcms\ for striations, while for strands it takes intermediate values $ N(\mathrm{H_2}) \approx (2-5) \times 10^{21} $~\pcms\ \citep{Cox16}.

The radial filament density profile is well described by the Plummer function:
\begin{equation}
\rho(r)=\frac{\rho_c}{\left[ 1+\left( r/R_0\right) ^2\right] ^{p/2} } \; ,
\label{eq:rho}
\end{equation}
where $ \rho_c $ is the central density and $ R_0 $ is the radius of the central flat part.
Such a profile is obtained by solving the problem of hydrostatic equilibrium of an infinite self-gravitating isothermal cylinder \citep{Stodolkiewicz63, Ostriker64}.
The theoretical value of $ p $ for an isothermal cylinder is 4. At the same time, observations give values close to 2 \citep{Andre14}. The reason for the discrepancies may be the non-isothermal nature of the filaments. The value of $ R_0 $ is
\begin{equation}
R_0=\frac{c_s^2}{G\Sigma_0} \; ,
\label{eq:r0}
\end{equation}
where $ c_s $ is the speed of sound, $ G $ is the gravitational constant, $ \Sigma_0 $ is the surface density \citep{Larson85, Hartmann02}.

The gravitational stability of filaments is determined by the amount of mass per unit length. The critical value of this parameter is
\begin{equation}
M_\mathrm{line,c} =\frac{2c_s^2}{G} \; .
\label{eq:mcrit}
\end{equation}
Taking into account non-thermal motions, the speed of sound should be replaced by $ \sigma_\mathrm{tot} =\sqrt{c_s^2 + \sigma_\mathrm{NT}^2} $ \citep{Fiege00}.
Filaments whose mass per unit length exceeds the critical value fragment and collapse. Jeans length along the axis is \citep{Larson85, Hartmann02}
\begin{equation}
\lambda_c=3.94\frac{c_s^2}{G\Sigma_0} \; .
\label{eq:l_c}
\end{equation}

For a more realistic description of filaments, it is also necessary to take into account the external pressure \citep{Fischera12}. In some cases, the observations are well described by the model of a pressure-limited filament surrounded by a low-density shell \citep{Kainulainen16}.

It was noted above that when the critical value of mass per unit length is exceeded, the process of its fragmentation can begin to develop in the filament. Such a phenomenon is actually observed in interstellar filaments at different scales \citep[for example,][]{Andre10, Offner14, Sanchez-Monge14, Beuther15, Samal15, Kainulainen17, Lu18, Ryabukhina18}. In some cases, there are signs of longitudinal collapse of the substance in the filaments \citep{Hacar13, Peretto14, Hacar17, Kirsanova17, Ryabukhina18, Ryabukhina21}. Fragmentation of filaments may be the key to understanding the initial mass function of stars, since it allows explaining the position of the peak and the shape of this function \citep{Andre10, Andre17}. 

\begin{figure}[htb]
    \centering
    \includegraphics[width=\textwidth]{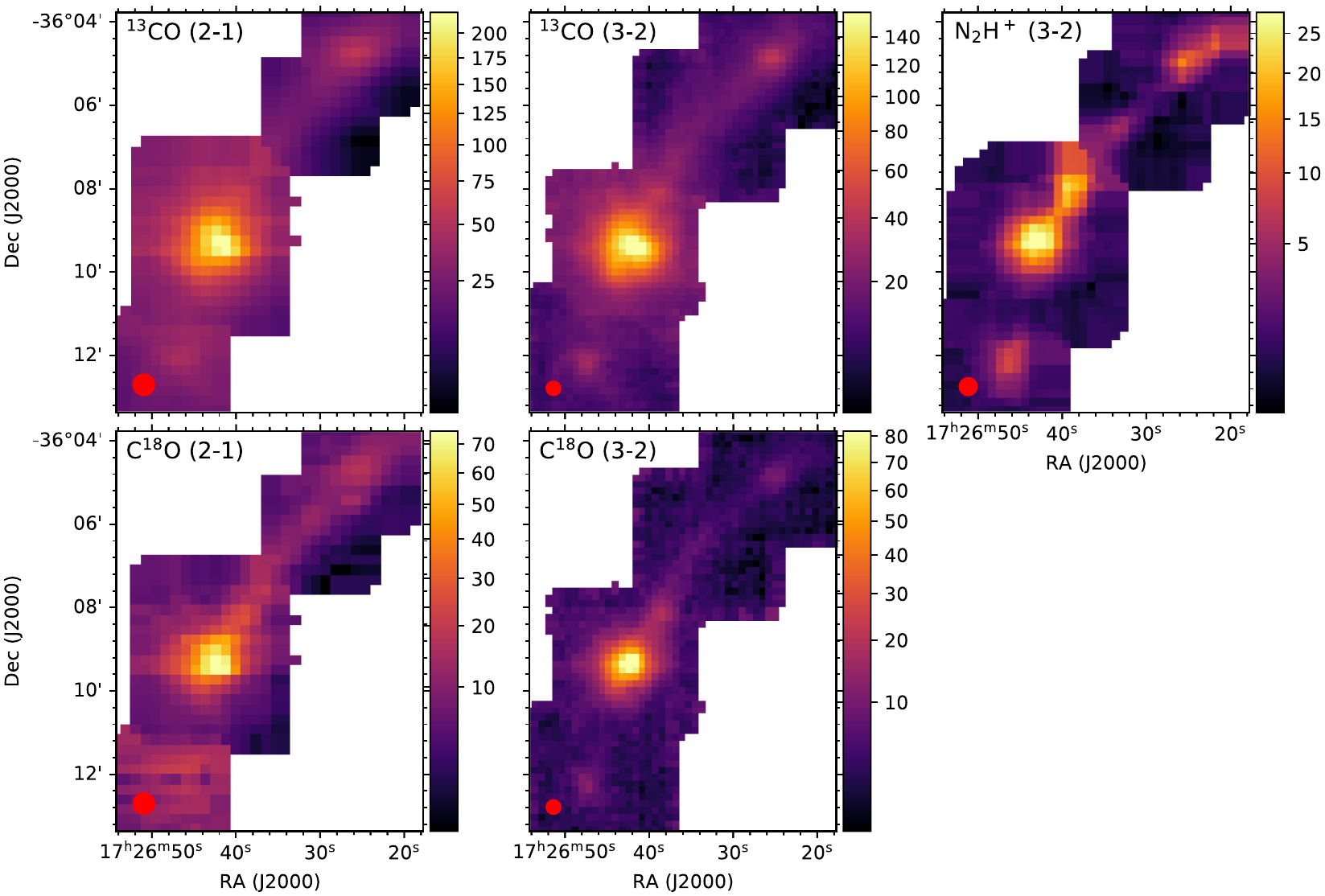}
    \caption{Integrated intensity maps of IRDC G351.78--0.54 in the lines $^{13}$CO (2--1), $^{13}$CO (3--2), C$^{18}$O (2--1), C$^{18}$O (3--2), and N$_2$H$^+$ (3--2). The red circles in the lower left corners indicate the beam size (HPBW). Adopted from \citet{Ryabukhina21}.}
    \label{fig:g351_maps}
\end{figure}

An example of our study of the filamentous IRDC G351.78--0.54 is shown in Fig.~\ref{fig:g351_maps} \citep{Ryabukhina21}. The total mass of the filament is estimated at $\sim 1800$~M\sun. The mass per unit length ($M_{\rm line}$ = 529 M$_\odot$/pc) is close to the critical value. However, both values represent upper limits. The presence of several dense clumps along the filament shows that the process of fragmentation is going on. Six dense clumps are identified in the N$_2$H$^+$ (3--2) map. All clumps except one appear gravitationally unstable.

Observations show that in some cases the process of star formation proceeds faster at the ends of filaments \citep{Beuther15, Kainulainen16, Dewangan19}, which may be due to the acceleration of matter in these regions \citep{Clarke15}.

In many cases, the most active star formation is observed at the intersections of the filaments. There are indications that this process may be initiated by filament collision \citep[e.g.][]{Nakamura14, Fukui15, Dewangan17}. In addition, \citet{Myers09} drew attention to the fact that many star-forming regions have a structure that can be described as a ``hub'' (an object with an increased radial density of matter and an axes ratio close to unity) and filaments. In this work, a model was proposed for the formation of such structures, in which it develops from an initially inhomogeneous medium of low density, which is compressed into a self-gravitating layer under the influence of external factors. The filamentous structure arises under the action of shock waves associated with compression or gravitational instabilities.

In general, numerous studies show that filaments naturally arise as a result of supersonic turbulence and shock waves  \citep[e.g.][]{Andre14, Inoue18}. They can also arise as a result of fragmentation of planar structures (for example, shells around \Hii\ zones, old supernova remnants, etc.).
A number of papers consider the mechanisms of formation of ``striations'' adjacent to filaments. A recent analysis of several possible mechanisms has shown that the most probable is the nonlinear interaction of MHD waves \citep{Tritsis16}.

\section{General structure of HMSF regions} \label{sec:sgtructure}
The general structure of HMSF regions is usually very complicated. They may contain many dense cores, filamentary structures, \Hii\ regions, outflows. There are numerous investigations of such regions. Here we briefly describe several regions covered by our studies.

\subsection{S187}
\begin{figure}
    \begin{minipage}{0.49\textwidth}
    \centering
	\includegraphics[width=\linewidth]{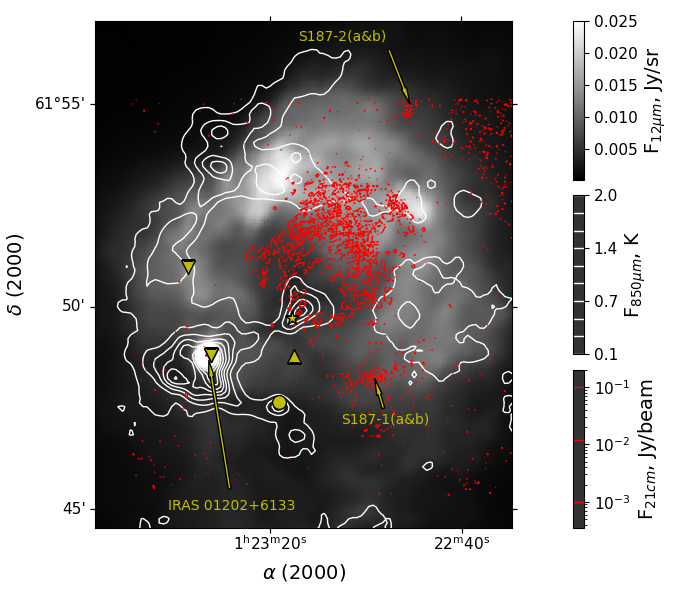}
    \end{minipage}
    \hfill
    \begin{minipage}{0.49\textwidth}
    \centering
	\includegraphics[width=\linewidth]{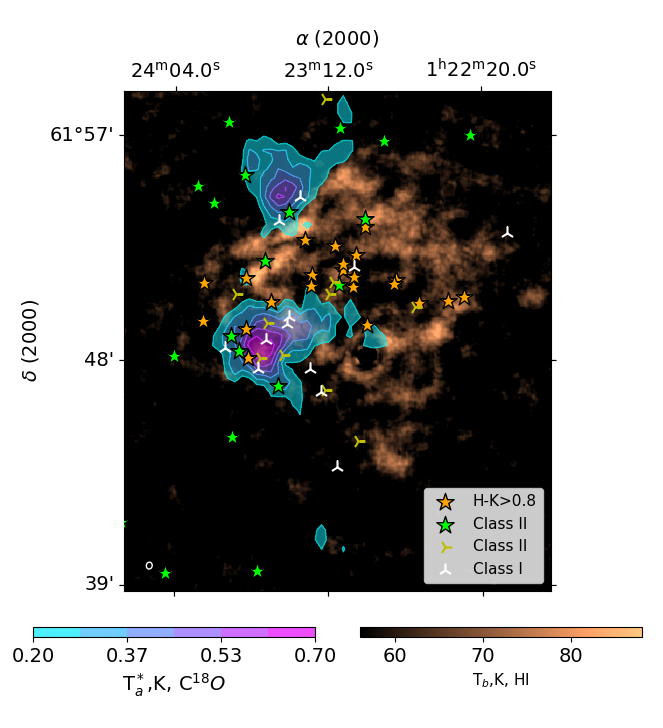}
    \end{minipage}
	\caption{Left panel: the structure of the S187  {complex}. The 1420 MHz GMRT continuum is in red {contours}, tracing the ionized gas. Bright red {spots} are related to radio galaxies (S187-1(a\&b) and S187-2(a\&b)){. The red contours near S187-1(a\&b) are caused by limited dynamical range. }
The 12$\mu$m WSSA image is {shown in} grayscale  tracing the dust in shocked gas. 
	White contours represent SCUBA~850$\mu m$ emission tracing cold dust in molecular material. 
	The water maser is marked as a triangle and the {OH masers from \citet{Engels15}} are marked as reversed triangles. {S187~NIRS~1 is shown as {a} circle {and S187 H$\alpha$ as the star}.} 
	Right panel: The distribution of the YSOs near S187. The background image represents the \Hi\ averaged emission. The blue contours represent the C$^{18}$O emission.
	Adopted from \citet{Zemlyanukha22}.}
	\label{fig:s187}
\end{figure}

The S187 \Hii\ region (Fig.~\ref{fig:s187}) at a distance of 1.4$\pm$0.26~kpc from photometry \citep{Russeil07} or $\sim$0.9~kpc from Gaia~DR2 data is surrounded by a molecular and atomic gas shell \citep{Joncas92, Arvidsson11}. A number of young stellar objects and molecular masers were detected towards the shell \citep{Kang17, Richards12, Engels15, Valdettaro01}, indicating ongoing star formation. The 1.2~mm continuum map and data on C$^{18}$O, CS, C$^{34}$S, HCN, H$^{13}$CN, HNC, HN$^{13}$C, HCO$^+$, H$^{13}$CO$^+$, N$_2$H$^+$ molecular emission are presented by \citet{Zin09}.

Recently we performed a detailed high resolution study of this region in the \Hi\ line at 21~cm with the Indian Giant Metrewave Radio Telescope (GMRT) \citep{Zemlyanukha22}. The achieved angular resolution is 8~arcsec (0.06~pc), which is the highest resolution available of the \Hi\ emission observations of Galactic regions. Such observations are challenging due to a large amount of atomic hydrogen on the line of sight. Main parameters (mass, spin temperature) of the atomic gas were determined. It was found that the atomic shell is highly inhomogeneous (Fig.~\ref{fig:s187}) and contains $\sim$100~fragments  {of} median mass around $\sim$1.1~M$_\odot$. The index of the mass-size power-law relationship is 2.33--2.6, which is close to values for molecular gas clumps and clouds. Two molecular cores at different stages of evolution are identified. The data indicate an interaction between the atomic shell and the molecular core.

\subsection{S255}
\begin{figure}
    \centering
    \includegraphics[width=\textwidth]{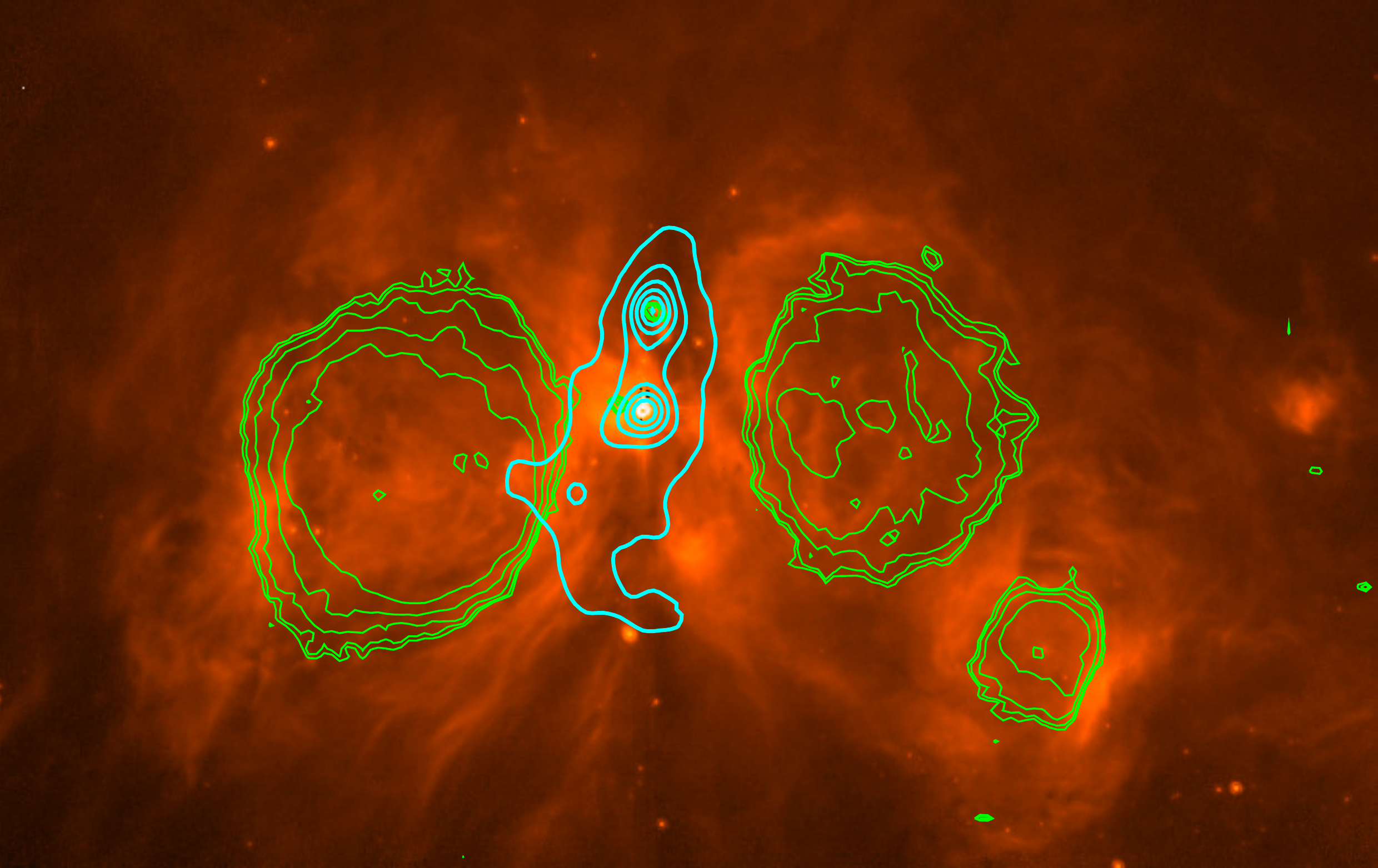}
    \caption{Maps of the S255 area at 610~MHz (thin green contours) and at 1.2~mm (thick cyan contours) overlaid on the 8~$\mu$m Spitzer image. Adopted from \citet{Zin17-sao}.}
    \label{fig:s255}
\end{figure}

The distance to the S255 star-forming region derived from the annual parallax measurements of water masers is $ 1.78^{+0.12}_{-0.11} $~kpc \citep{Burns16}. It looks as a ridge of the molecular and dust emission sandwiched between the two evolved \Hii\ regions S255 and S257 (Fig.~\ref{fig:s255}. 

The NIR observations show that the sources associated with the gas ridge are younger than the sources outside the gas ridge, which hints at induced star formation \citep{Ojha11}. This ridge contains two major clumps, S255IR and S255N \citep[e.g.][]{Wang11, Zin12}. Both of these clumps represent sites of massive star formation and contain several dense cores. Detailed studies of these clumps are presented in \citet{Zin15, Zin18-raa, Zin18-iau, Zin20, Liu18, Liu20, Zemlyanukha18}.

In S255IR the ALMA observations reveal a very narrow ($\sim$1000--1800~AU) and dense ($n\sim 3\times 10^7$~\pcmc\ assuming the cylindrical geometry) filamentary structure {with} at least two velocity components \citep{Zin20}. The mass estimated from the continuum emission is about 35~M$_\odot$. Two star-forming cores are apparently associated with this structure. There are molecular outflows originating in these cores (see Sect.~\ref{sec:disks_outflows}). The most massive core, S255IR-SMA1, contains a 20~M\sun\ protostar, which shows signs of episodic accretion. It is surrounded by a rotating and infalling envelope. Several new low-mass prestellar cores are discovered.

The S255N clump contains several large fragments at different velocities \citep{Zemlyanukha18}. The central core, S255N-SMA1, is resolved into two components with significantly different temperatures ($\sim$150~K and $\sim$25~K). The bipolar outflow is associated with the hot source. It is surrounded by a large torus (Sect.~\ref{sec:disks_outflows}). There are several other cores with very young outflows in this area.

\subsection{W40} \label{sec:w40}
\begin{figure}
    \centering
    \begin{minipage}[b]{0.522\textwidth}
    \includegraphics[angle=-90, width=\textwidth]{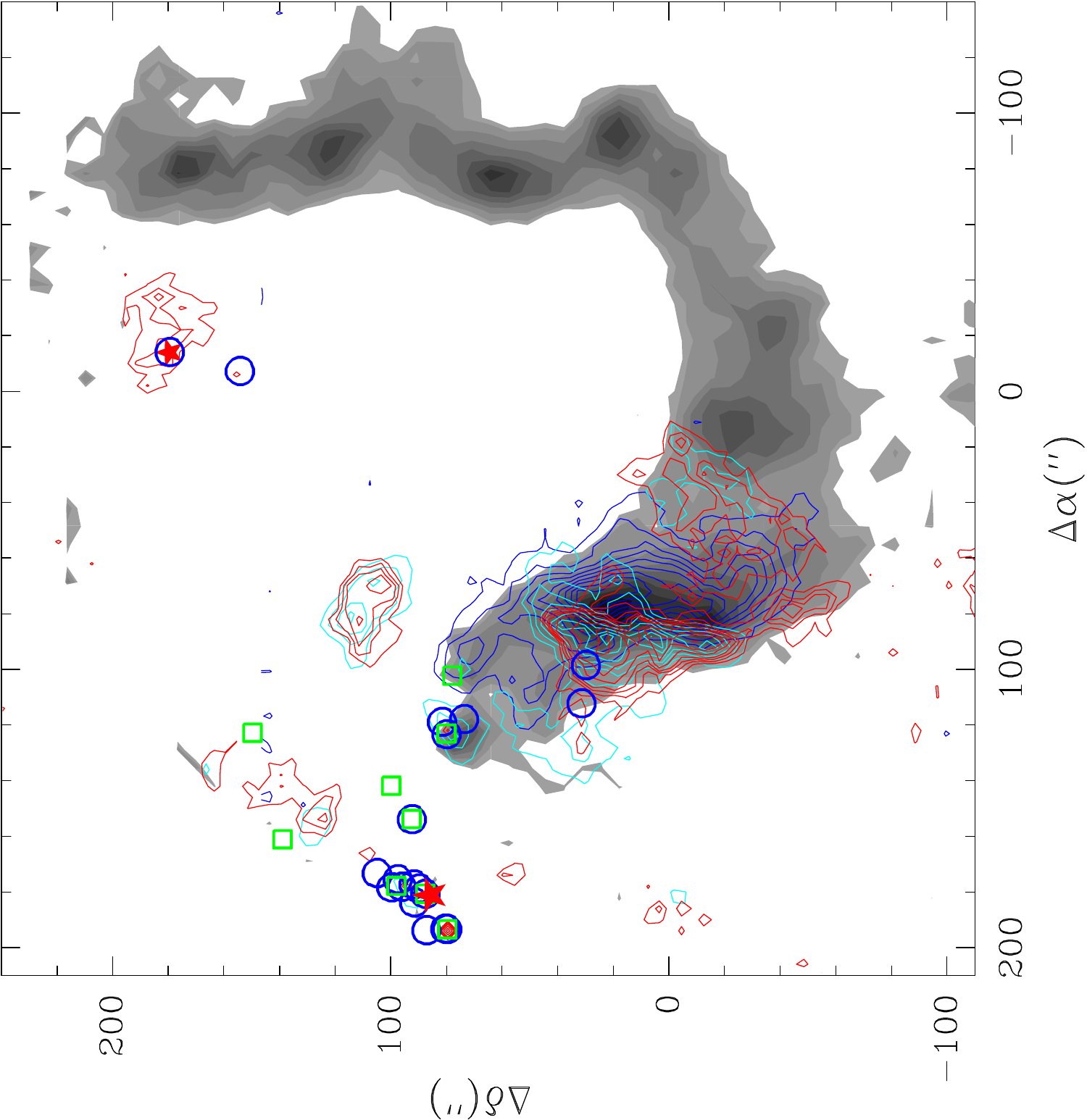}
    \end{minipage}
    \hfill
    \begin{minipage}[b]{0.468\textwidth}
    \includegraphics[angle=-90, width=\textwidth]{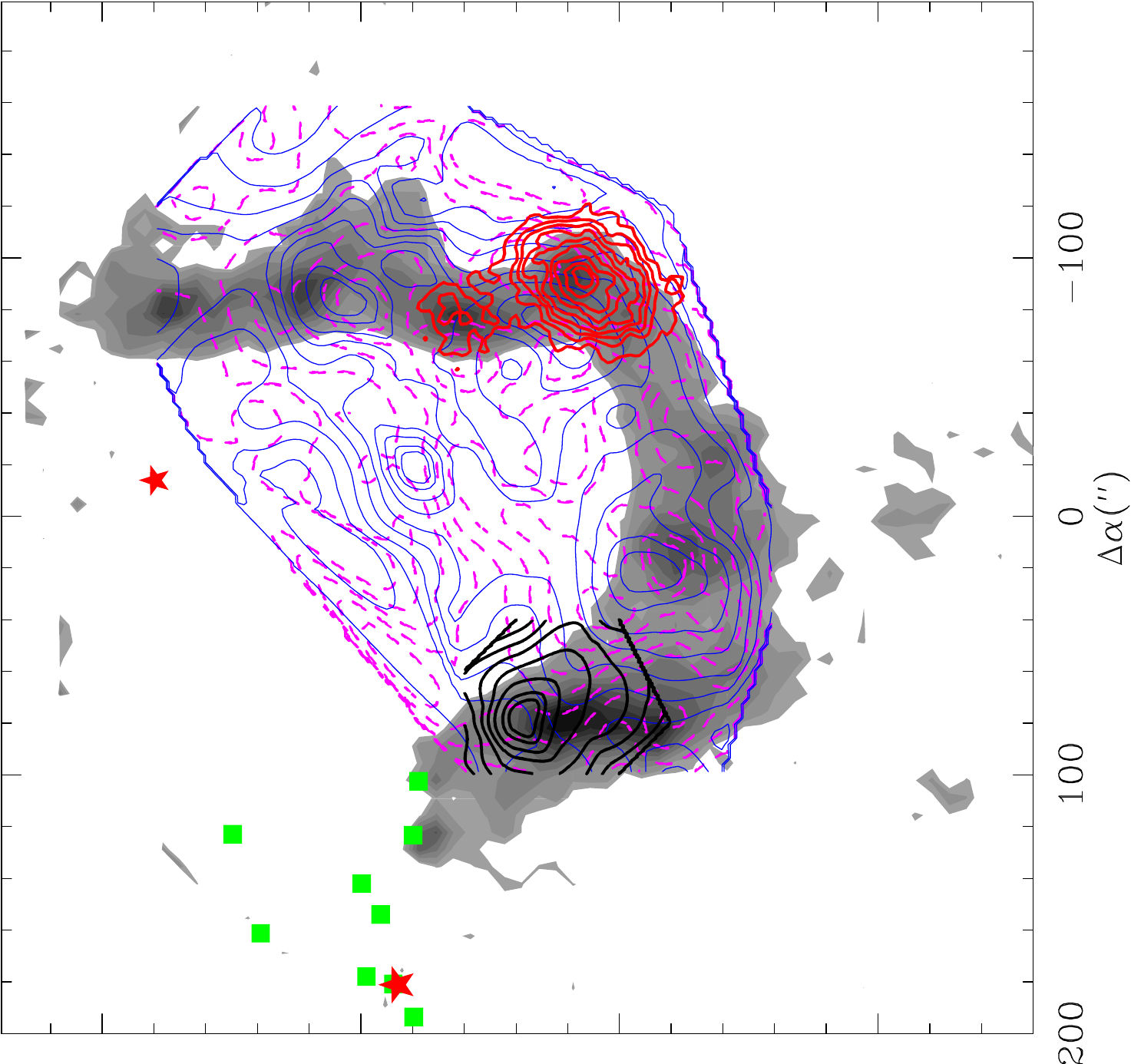}
    \end{minipage}
    \caption{Left panel: the high-resolution 1280 and 610 MHz maps (spatial resolutions $\sim$2.4 and $\sim$5 arcsec, respectively) and the CS(5--4) integrated intensity map (red, cyan and blue contours, respectively). Right panel: Molecular-line integrated intensity maps overlaid on1.2-mm dust continuum emission (grey-scale). HCN(1--0) (crimson dashed contours), HCO$^+$(1--0) (blue contours), N$_2$H$^+$(1--0) (red contours) and C$^{34}$S(2--1) (black contours). The figures are from \citet{Pirogov13} where the other symbols are explained.}
    \label{fig:w40}
\end{figure}

The nearby \Hii\ region W40 was investigated in detail by \citet{Mallick13, Pirogov13}. The data include the dust continuum, line molecular emission and observations of the ionized gas. A clumpy dust ring is revealed (Fig.~\ref{fig:w40}), which is probably formed by the collect-and-collapse process due to expansion of the neighbouring \Hii\ region. Nine dust clumps in the ring have been identified. The distributions of the dust and molecular emission are very different and a strong molecular chemical differentiation is observed. Molecular and electron abundances in different parts of the source are estimated. There are signs of triggered star formation in a part of the ring.


\subsection{W42}
The W42 region at a distance of 3.8~kpc contains a rare O-type protostar W42-MME (mass: 19$\pm$4~M\sun). The area around this protostar was investigated in various molecular lines and in continuum with the Atacama Large Millimeter/ submillimeter Array (ALMA), Submillimeter Array, and Very Large Array at a high angular resolution of $\sim$0{\farcs}3--3{\farcs}5 \citep{Dewangan22}.

The data show no \Hii\ zone around W42-MME, which confirms its protostellar nature. There is an elongated structure with a signature of Keplerian rotation in the center, associated with the bipolar molecular outflow observed in the CO and SiO lines. However the angular resolution is insufficient to resolve the probable disk. Several knots of the SiO emission hint at episodic ejections. The temperature estimates from the CH$_3$CN observations ($\sim$220~K) indicate the presence of a hot core. The core hosting W42-MME appears to gain mass from the envelope and also from the immediate surrounding cores.

\section{Prestellar and protostellar cores, disks and bipolar outflows} \label{sec:cores}

\subsection{Properties of dense cores} 
Ultimately, the process of star formation occurs in the so-called dense cores, which can be formed, for example, as a result of the fragmentation of interstellar filaments. In Section~\ref{sec:surveys} we described the search and study of such cores. At present, a certain classification of such cores has been developed, based on the data on the presence of (proto)stellar objects in them and on their spectral characteristics. It has been developed in most detail for cores in low-mass star formation regions. They are usually divided into several main categories: starless, prestellar, protostellar \citep{diFrancesco07}. Prestellar, unlike stellarless ones, are gravitationally bound and can later form a protostar. In protostellar, such a protostar already exists. For protostellar cores, there is a well-established division into 4 classes - from Class~0 to Class~III, based on their spectral characteristics and obviously corresponding to the evolutionary sequence \citep{Lada87, Andre93, Andre00}.

A lot of effort has gone into searching for massive prestellar cores (with a mass of $ \sim 30 $~M$_\odot$ within a radius of $ 0.03 $~pc) that could form a massive protostar. The discovery of such cores could serve as an argument in support of the monolithic collapse model during the formation of massive stars. So far, only a few candidates for such cores have been discovered \citep{Louvet18}. The absence of a noticeable number of massive prestellar cores can be explained by their short lifetime, on the order of free fall time.

Many works are devoted to the study of statistical distributions of the parameters of cores. In particular, the mass distribution of low-mass prestellar cores is very similar to the initial mass function of stars \citep{Andre10}. This indicates an approximately constant star formation efficiency across the mass spectrum. The core mass function itself can be explained by filament fragmentation, taking into account turbulence \citep{Andre17}.

Of great importance for understanding the processes in interstellar clouds are the so-called Larson's laws, obtained by him from the analysis of observational data in the early 80s \citep{Larson81}. The first one is the relation between velocity dispersion and size, $ \sigma \propto L^\alpha $, where $ \alpha \approx 0.5 $. The second law says that clouds are gravitationally bound, the virial parameter $ \alpha_\mathrm{vir} \sim 1 $. According to the third law, the column density in clouds is approximately the same. These relations are not independent, each of them follows from the other two. These ratios are most likely a consequence of turbulence in interstellar clouds.

Larson's laws were originally derived for giant molecular clouds and their applicability to dense clumps was investigated separately. It was found in \citet{DCDCL7, Caselli95} that the Larson relation between nonthermal velocity dispersion and size is well satisfied for dense cores in dark clouds.

As for massive cores, the results here are rather contradictory. \citet{Caselli95} derived the exponent for $ (\sigma - L) $ $ \alpha \approx 0.2 $. \citet{Plume97} did not find any relationship between velocity dispersion and size, while \citet{Pirogov98} obtained the exponent $\alpha \approx 0.5 $.

The radial density profile for cores in dark clouds is well described by the Bonnor-Ebert model, with starless cores close to the Bonnor-Ebert critical sphere, and protostellar cores corresponding to the supercritical case \citep{Alves01, Kirk05, McKee07}. Similar results were also obtained for massive cores \citep{Pirogov09}.

The radial gas temperature profile for several massive cores was investigated by \citet{Zin05, Malafeev06} on the basis of CH$_3$CCH observations. They found that it can be fitted by a power law with --0.3 ... --0.4 indices. This dependence is in agreement with the theoretically expected one for a centrally heated optically thin cloud. The dust temperature profiles were analyzed recently by \citet{Pirogov22x}. He found that they are also consistent with heating by the central source.

\subsection{Observations of disks and bipolar outflows} \label{sec:disks_outflows}
At present, it is generally accepted that the formation of stars with a mass of the order of the sun occurs by disk accretion of matter, accompanied by bipolar outflows. There are observations of a large number of disks in low mass star formation regions \citep[e.g.][]{Andrews05, Andrews07}. There are also observations of disk candidates or toroidal structures around several dozen massive protostars \citep{Beltran17}, although convincing cases are still very rare. High-velocity bipolar outflows are observed everywhere. A detailed review of their characteristics and models is presented in \citet{Arce07}.

Of particular interest are observations of disks and outflows in HMSF regions, since they allow a better understanding of the mechanism of formation of such stars. This mechanism is still unclear and is actively debated  \citep[e.g.][]{McKee07, Tan14}. The main models discussed are the monolithic collapse of a massive dense core and competitive accretion. Quite exotic models are proposed, in which massive stars are formed by merging stars of smaller mass \citep{Bonnell98}.

Recently, several events have been recorded that support disk accretion as a mechanism for the formation of stars with masses up to at least $ \sim$20~M$_\odot$. The first two are outbursts of the luminosity of objects NGC6334I-MM1 \citep{Hunter17} and S255IR NIRS3/SMA1 \citep{Caratti16, Liu18}. They were accompanied by maser bursts \citep{Fujisawa15, Moscadelli17, Zin17, Hunter18}. These events have been interpreted as the result of episodic accretion of matter onto a central massive protostar, similar to that observed in the formation of low-mass stars, but on a much larger scale. This roughly corresponds to some theoretical models of fragmented disks around massive protostars \citep{Meyer17}.

Another example of a disk-outflow system around a massive YSO, which shows signs of episodic accretion, is the O-type protostar W42-MME \citep{Dewangan22}.

\begin{figure}
    \centering
    \includegraphics[angle=-90,width=0.8\textwidth]{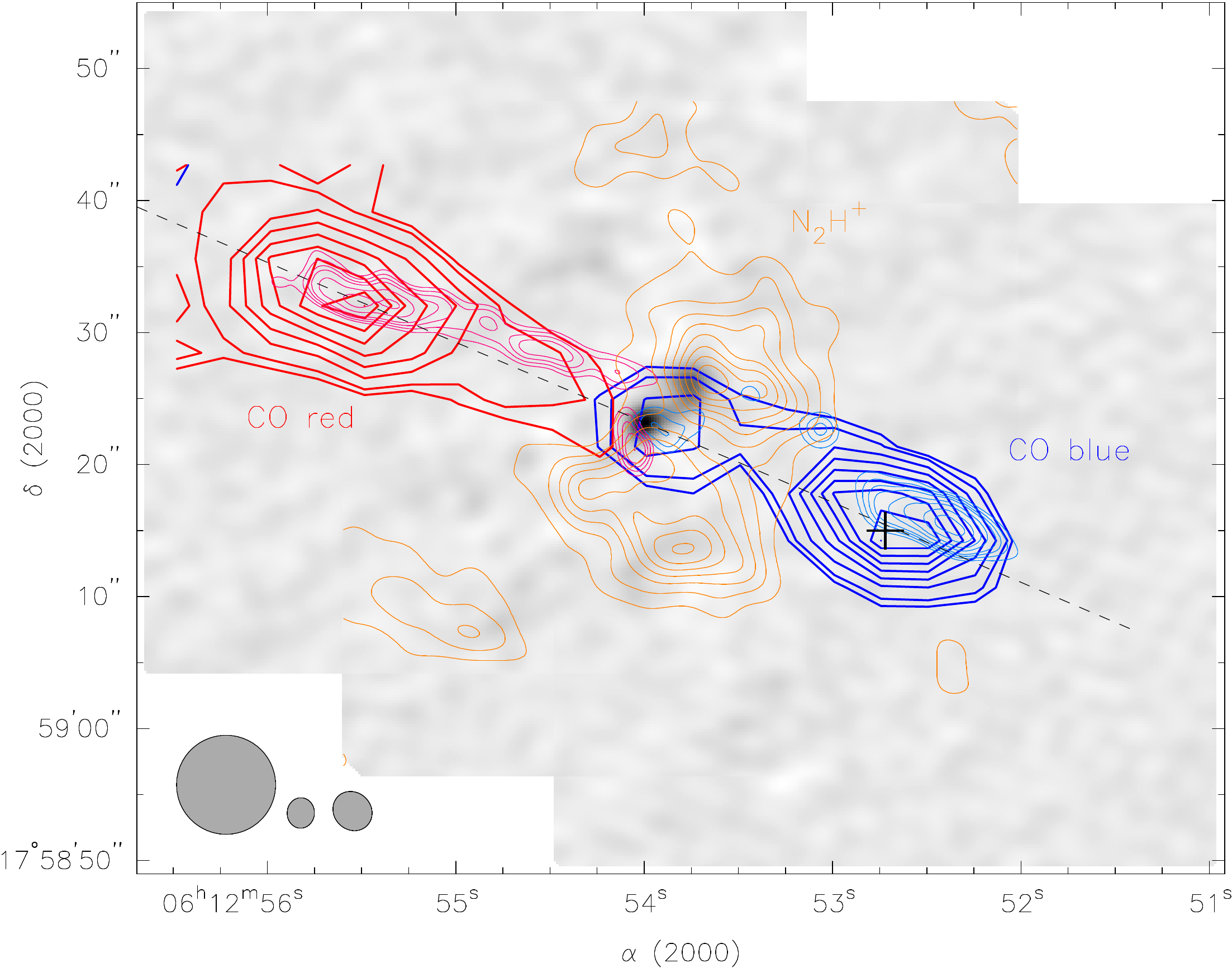}
    \caption{Maps of the CO(3--2) high velocity emission as observed with the IRAM-30m telescope (blue and red thick contours) in the S255IR area overlaid on the continuum image at 0.8~mm. The thin contours show the SMA maps. The dashed line indicates the axis of the jet (P.A. = 67°). The orange contours show the N$_2$H$^+$(3--2) integrated line emission obtained by combining the SMA and 30m data. The cross marks the position of the high-velocity dense clump. The 30m beam and the SMA beams for the CO and N$_2$H$^+$ observations, respectively, are shown in the lower left corner (from left to right) Adopted from \citet{Zin15}.
}
    \label{fig:s255ir-outflow}
\end{figure}

For S255IR NIRS3/SMA1, modeling with radiative transfer calculations the kinematics of the CH$_3$CN gas shows that the CH$_3$CN emission is best described by a flattened rotating envelope with infalling motion \citep{Liu20}. A mass infall rate of a few $\times 10^{-4}$~M\sun\,yr$^{-1}$ is derived. This object drives a fast collimated jet and a molecular outflow with a wide opening angle, surrounded by dense walls observed in the C$^{34}$S line \citep{Zin20}. There is another outflow from the nearby SMA2 core. The large-scale view of these outflows is presented in Fig.~\ref{fig:s255ir-outflow}.

A combination of a fast jet and a slower outflow has been rarely observed in HMSF regions \citep[e.g.,][]{Torrelles11}, while it is known in low-mass stars \citep[e.g.,][]{Arce07} and theoretical models of such event have been developed. In addition to S255IR NIRS3/SMA1, another example of such combination was discovered recently in MYSO G18.88MME \citep{Zinchenko21}. The gas density in the fast component is very high since a rather strong HC$_3$N $J=24-23$ emission is observed in the high velocity gas. The critical density for this transition is $\sim 3\times 10^6$~\pcmc. Both components seem to be rotating and, if this interpretation of the transverse velocity gradient is correct, the specific angular momentum for the fast component is very high, $\sim 10^4$~AU\,\kms.

Rotating structures around very massive stars may represent toroids \citep[e.g.][]{Cesaroni07, Beltran11, Beltran16}. A large torus with the inner radius of 8000~AU and the outer radius of 12000~AU was identified in the S255N region by \citet{Zemlyanukha18}. They argue that it plays an important role in the process of mass accumulation by the central protostar.

\section{Chemistry in high-mass star-forming regions} \label{sec:chemistry}

\subsection{Chemical differentiation}
Maps of star-forming regions (and HMSF regions in particular) in lines of different molecules are frequently very different. Although these differences can be partly caused by different excitation of these molecules and opacity effects, in most cases they reflect variations of relative molecular abundances. This molecular chemical differentiation can be related to variations of the physical parameters and to evolution of the chemical content with time. Examples of such differentiation in our observations are presented in \citet{Zin05, Zin09, Zin11-iau, Zin18-iau, Zinchenko22, Lintott05, Pirogov07, Pirogov13}. 

In particular, it was found that the CS emission correlates well with the dust continuum emission, while the N$_2$H$^+$ intensity drops towards the CS peaks (associated with luminous IR sources) for most of the HMSF regions, which can be due to an N$_2$H$^+$ abundance decrease \citep{Pirogov07}. A drop of the N$_2$H$^+$ abundance towards the luminous IRAS source in the filamentary IRDC G351.78--0.54 was reported recently by \citet{Ryabukhina21}. \citet{Zin09} found that the abundances of CO, CS and HCN are more or less constant, while the abundances of HCO$^+$, HNC and especially N$_2$H$^+$ strongly vary in HMSF regions. They anticorrelate with the ionization fraction and as a result decrease towards the embedded YSOs. For N$_2$H$^+$ this can be explained by dissociative recombination to be the dominant destroying process.

A strong chemical differentiation in the W40 region (Sect.~\ref{sec:w40}) was reported by \citet{Pirogov13}. The CS abundance is enhanced towards the eastern dust clump 2, while the NH$_3$, N$_2$H$^+$ and H$^{13}$CO$^+$ abundances are enhanced towards the western clumps. HCN and HCO$^+$ do not correlate with the dust, probably tracing the surrounding gas. 

\subsection{Deuteration}

The effect of deuterium fractionation in dense interstellar clouds (i.e., an increase of the relative abundance of deuterated molecules) has been investigated for a long time already. It is explained by the exothermicity of the reactions of replacing a proton with deuteron in molecules, which underlie the chains of the chemical reactions leading to the formation of most other molecules, at first H$_3^+$ \citep[e.g.][]{Roueff07}. In addition, at low temperatures, freezing of molecules which destroy H$_2$D$^+$, in particular CO, on dust grains is important, as well as the decreased ionization degree which reduces the recombination rate of H$_2$D$^+$.

This effect is temperature dependent and has been mostly investigated in cold clouds. however recent studies show that it is rather efficient in warmer clouds, where massive stars form \citep[e.g.][]{Gerner15}. In order to investigate it in HMSF regions in more detail we performed a survey of about 50 such regions in the lines of deuterated molecules at 3--4~mm with the 20-m Onsala radio telescope \citep{Trofimova20}. The $J=1-0$ transitions of DCN, DNC, DCO$^+$, N$_2$D$^+$ and the ortho-NH$_2$D $1_{11}-1_{01}$ line were observed. DCO$^+$,  DCN, DNC and NH$_2$D were detected in about 1/3 of the observed sources, while N$_2$D$^+$ was only seen in two. The dependencies of the abundances of these molecules on temperature and velocity dispersion were analyzed taking into account data represented by upper and lower limits. A statistically significant decrease of the DCO$^+$ abundance with increasing temperature was found, while the DCN abundance remains nearly constant. There is a noticeable decrease of the DCO$^+$ abundance with increasing line width. The DCN/HCN ratio is $\sim$10$^{-2}$ for the sources detected in the DCN line and remains nearly constant in the temperature range $15-50$~K. The NH$_2$D/NH$_3$ ratio also remains practically constant ($\sim$10$^{-2}$) in this temperature range (Trofimova et al., in preparation), which contradicts the chemical model for ammonia deuteration presented in \cite{Roueff05}, predicting its drop to $\sim 2\times 10^{-3}$ at 50~K.

\begin{figure}[h]
\begin{minipage}[t]{0.325\linewidth}
\center{\includegraphics[width=\linewidth]{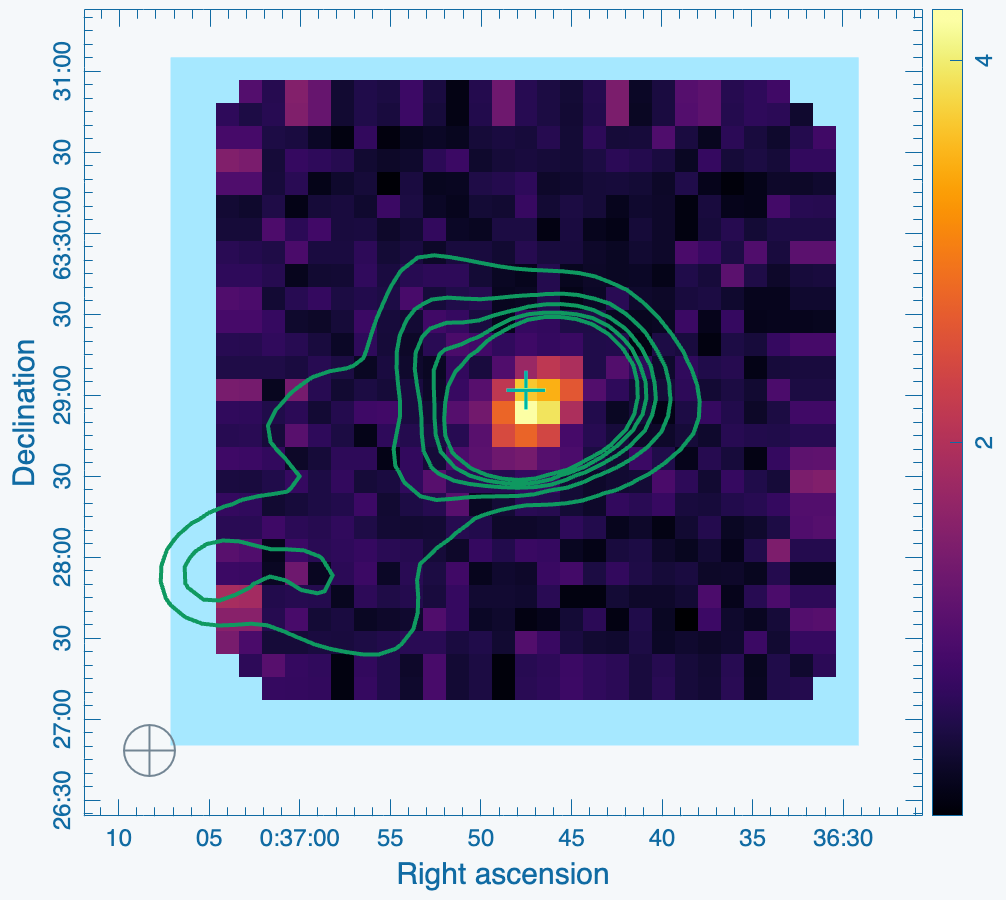}} 
\end{minipage}
\hfill
\begin{minipage}[t]{0.325\linewidth}
\center{\includegraphics[width=\linewidth]{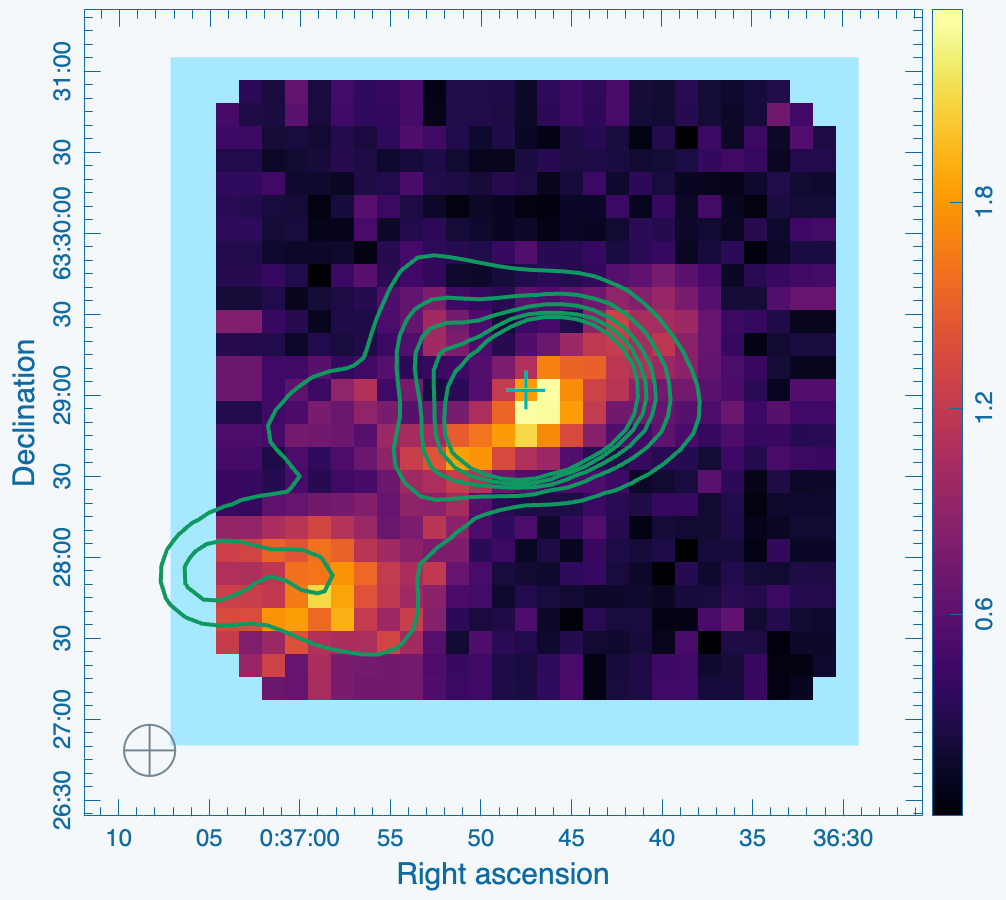}} 
\end{minipage}
\hfill
\begin{minipage}[t]{0.325\linewidth}
\center{\includegraphics[width=\linewidth]{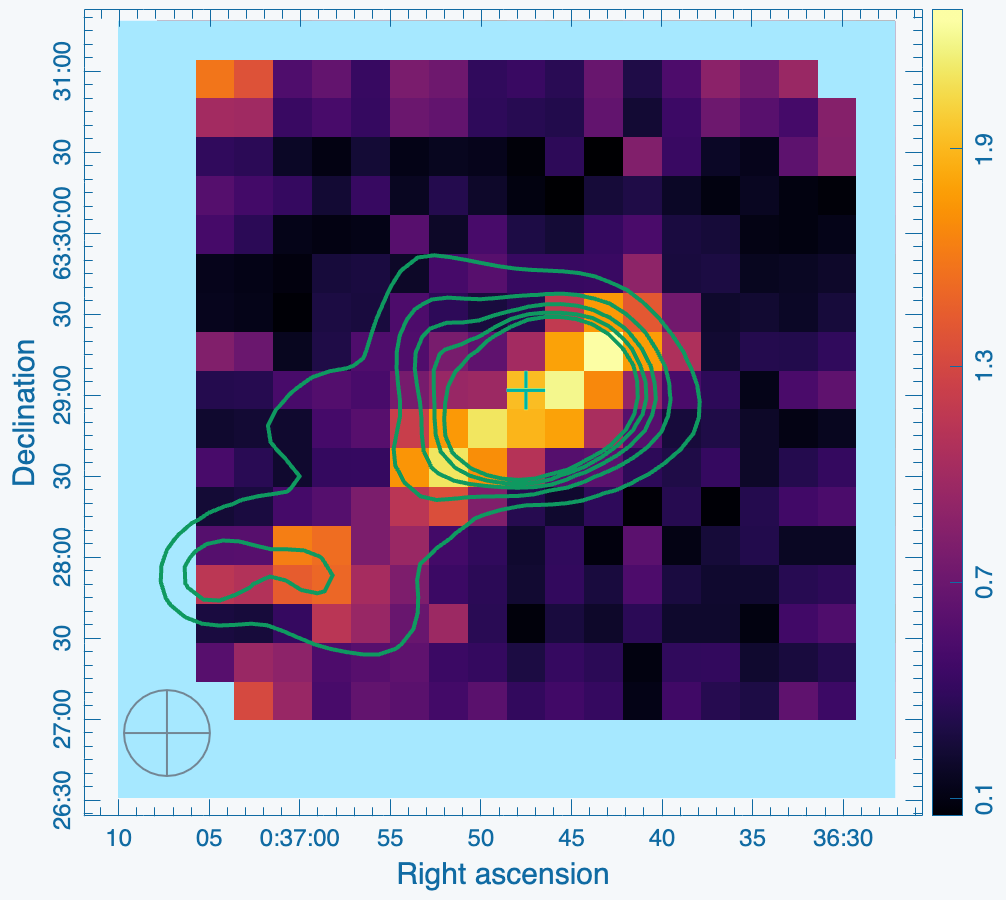}} 
\end{minipage}
\caption{Maps of L1287 in the DCN, DCO$^+$ $J=2-1$ and NH$_2$D $1_{11}-1_{01}$ lines (from left to right) obtained with the 30-m IRAM radio telescope. The contours represent the H$_2$ column density. The plus sign indicates the position of the IRAS source. The telescope beam is indicated in the lower left corner. Adopted from \citet{Zinchenko22}.
    }
\label{fig:l1287-maps}
\end{figure}

Later several sources were mapped in these lines and in the lines of higher transitions of these molecules at 2 mm with the 30-m IRAM radio telescope \citep{Zinchenko22}. These maps show significant differences between distributions of the targeted molecules (see an example in Fig.~\ref{fig:l1287-maps}). 

In general, the DCN peaks are observed towards the temperature peaks, which coincide with locations of luminous IR sources, while other deuterated molecules trace colder regions. We derived the gas volume density and molecular column densities from the $J=1-0$ and $J=2-1$ transitions of DCN, DNC and DCO$^+$ by non-LTE modeling with RADEX. Then, deuteration degrees for these species were obtained. The deuteration degrees for HCO$^+$ and HNC drop with increasing temperature, while for HCN it is more or less constant.

\section{Conclusions}
Studies of high mass star formation remain a hot topic of astrophysical research. There are many open questions in this area and the general scenario of this process is still under debate. New facilities provide an inflow of new important data, which creates good prospects for new discoveries. 

This research was partly supported by the IAP RAS state program 0030-2021-0005 and by the Russian Science Foundation grant No. 22-22-00809 (Sect.~\ref{sec:chemistry}).

\bibliography{sf}

\begin{thebibliography}{161}
\providecommand{\natexlab}[1]{#1}

\bibitem[\protect\astroncite{{Aguirre} et~al.}{2011}]{Aguirre11}
{Aguirre} J.E., {Ginsburg} A.G., {Dunham} M.K., et~al. (2011) \apjs 192, 4

\bibitem[\protect\astroncite{{Alves} et~al.}{2001}]{Alves01}
{Alves} J.F., {Lada} C.J., {Lada} E.A. (2001) \nat 409, 159

\bibitem[\protect\astroncite{{Andr{\'e}} et~al.}{2014}]{Andre14}
{Andr{\'e}} P., {Di Francesco} J., {Ward-Thompson} D., et~al. (2014) In:
  Protostars and Planets VI, eds. H.~{Beuther}, R.S. {Klessen}, C.P.
  {Dullemond}, T.~{Henning}. 27

\bibitem[\protect\astroncite{{Andr{\'e}} et~al.}{2017}]{Andre17}
{Andr{\'e}} P., {K{\"o}nyves} V., {Arzoumanian} D., et~al. (2017) \memsai 88,
  521

\bibitem[\protect\astroncite{{Andr{\'e}} et~al.}{2010}]{Andre10}
{Andr{\'e}} P., {Men'shchikov} A., {Bontemps} S., et~al. (2010) \aap 518, L102

\bibitem[\protect\astroncite{{Andre} et~al.}{1993}]{Andre93}
{Andre} P., {Ward-Thompson} D., {Barsony} M. (1993) \apj 406, 122

\bibitem[\protect\astroncite{{Andre} et~al.}{2000}]{Andre00}
{Andre} P., {Ward-Thompson} D., {Barsony} M. (2000) In: Protostars and Planets
  IV, eds. V.~{Mannings}, A.P. {Boss}, S.S. {Russell}. 59

\bibitem[\protect\astroncite{{Andrews} and {Williams}}{2005}]{Andrews05}
{Andrews} S.M., {Williams} J.P. (2005) \apj 631, 1134

\bibitem[\protect\astroncite{{Andrews} and {Williams}}{2007}]{Andrews07}
{Andrews} S.M., {Williams} J.P. (2007) \apj 659, 705

\bibitem[\protect\astroncite{{Arce} et~al.}{2007}]{Arce07}
{Arce} H.G., {Shepherd} D., {Gueth} F., et~al. (2007) In: Protostars and
  Planets V, eds. B.~{Reipurth}, D.~{Jewitt}, K.~{Keil}. 245

\bibitem[\protect\astroncite{{Arvidsson} and {Kerton}}{2011}]{Arvidsson11}
{Arvidsson} K., {Kerton} C.R. (2011) \aj 141, 153

\bibitem[\protect\astroncite{{Beltr{\'a}n} et~al.}{2011}]{Beltran11}
{Beltr{\'a}n} M.T., {Cesaroni} R., {Neri} R., et~al. (2011) \aap 525, A151

\bibitem[\protect\astroncite{{Beltr{\'a}n} and {de Wit}}{2016}]{Beltran16}
{Beltr{\'a}n} M.T., {de Wit} W.J. (2016) \aapr 24, 6

\bibitem[\protect\astroncite{{Beltr{\'a}n} and {de Wit}}{2017}]{Beltran17}
{Beltr{\'a}n} M.T., {de Wit} W.J. (2017) \memsai 88, 581

\bibitem[\protect\astroncite{{Benjamin} et~al.}{2003}]{Benjamin03}
{Benjamin} R.A., {Churchwell} E., {Babler} B.L., et~al. (2003) \pasp 115, 953

\bibitem[\protect\astroncite{{Benson} et~al.}{1998}]{DCDCL11}
{Benson} P.J., {Caselli} P., {Myers} P.C. (1998) \apj 506, 743

\bibitem[\protect\astroncite{{Benson} and {Myers}}{1983}]{DCDCL4}
{Benson} P.J., {Myers} P.C. (1983) \apj 270, 589

\bibitem[\protect\astroncite{{Benson} and {Myers}}{1989}]{Benson89}
{Benson} P.J., {Myers} P.C. (1989) \apjs 71, 89

\bibitem[\protect\astroncite{{Beuther} et~al.}{2015}]{Beuther15}
{Beuther} H., {Ragan} S.E., {Johnston} K., et~al. (2015) \aap 584, A67

\bibitem[\protect\astroncite{{Bonnell} et~al.}{1998}]{Bonnell98}
{Bonnell} I.A., {Bate} M.R., {Zinnecker} H. (1998) \mnras 298, 93

\bibitem[\protect\astroncite{{Brouillet} et~al.}{2005}]{Brouillet05}
{Brouillet} N., {Muller} S., {Herpin} F., et~al. (2005) \aap 429, 153

\bibitem[\protect\astroncite{{Burns} et~al.}{2016}]{Burns16}
{Burns} R.A., {Handa} T., {Nagayama} T., et~al. (2016) \mnras 460, 283

\bibitem[\protect\astroncite{{Burov} et~al.}{1988}]{Burov88}
{Burov} A.B., {Vdovin} V.F., {Zinchenko} I.I., et~al. (1988) Pisma v
  Astronomicheskii Zhurnal 14, 492

\bibitem[\protect\astroncite{{Caratti O Garatti} et~al.}{2017}]{Caratti16}
{Caratti O Garatti} A., {Stecklum} B., {Garcia Lopez} R., et~al. (2017) Nature
  Physics 13, 276

\bibitem[\protect\astroncite{{Carey} et~al.}{1998}]{Carey98}
{Carey} S.J., {Clark} F.O., {Egan} M.P., et~al. (1998) \apj 508, 721

\bibitem[\protect\astroncite{{Carey} et~al.}{2005}]{Carey05}
{Carey} S.J., {Noriega-Crespo} A., {Price} S.D., et~al. (2005) In: American
  Astronomical Society Meeting Abstracts, vol.~37 of Bulletin of the American
  Astronomical Society. 1252

\bibitem[\protect\astroncite{{Caselli} et~al.}{2002}]{DCDCL14}
{Caselli} P., {Benson} P.J., {Myers} P.C., et~al. (2002) \apj 572, 238

\bibitem[\protect\astroncite{{Caselli} and {Myers}}{1995}]{Caselli95}
{Caselli} P., {Myers} P.C. (1995) \apj 446, 665

\bibitem[\protect\astroncite{{Cesaroni} et~al.}{2007}]{Cesaroni07}
{Cesaroni} R., {Galli} D., {Lodato} G., et~al. (2007) In: Protostars and
  Planets V, eds. B.~{Reipurth}, D.~{Jewitt}, K.~{Keil}. 197

\bibitem[\protect\astroncite{{Clarke} and {Whitworth}}{2015}]{Clarke15}
{Clarke} S.D., {Whitworth} A.P. (2015) \mnras 449, 1819

\bibitem[\protect\astroncite{{Cox} et~al.}{2016}]{Cox16}
{Cox} N.L.J., {Arzoumanian} D., {Andr{\'e}} P., et~al. (2016) \aap 590, A110

\bibitem[\protect\astroncite{{Csengeri} et~al.}{2014}]{Csengeri14}
{Csengeri} T., {Urquhart} J.S., {Schuller} F., et~al. (2014) \aap 565, A75

\bibitem[\protect\astroncite{{Dame} et~al.}{2001}]{Dame01}
{Dame} T.M., {Hartmann} D., {Thaddeus} P. (2001) \apj 547, 792

\bibitem[\protect\astroncite{{Dewangan} et~al.}{2017}]{Dewangan17}
{Dewangan} L.K., {Ojha} D.K., {Zinchenko} I. (2017) \apj 851, 140

\bibitem[\protect\astroncite{{Dewangan} et~al.}{2019}]{Dewangan19}
{Dewangan} L.K., {Pirogov} L.E., {Ryabukhina} O.L., et~al. (2019) \apj 877, 1

\bibitem[\protect\astroncite{{Dewangan} et~al.}{2022}]{Dewangan22}
{Dewangan} L.K., {Zinchenko} I.I., {Zemlyanukha} P.M., et~al. (2022) \apj 925,
  41

\bibitem[\protect\astroncite{{di Francesco} et~al.}{2007}]{diFrancesco07}
{di Francesco} J., {Evans} N.~J. I., {Caselli} P., et~al. (2007) In: Protostars
  and Planets V, eds. B.~{Reipurth}, D.~{Jewitt}, K.~{Keil}. 17

\bibitem[\protect\astroncite{{Egan} et~al.}{1998}]{Egan98}
{Egan} M.P., {Shipman} R.F., {Price} S.D., et~al. (1998) \apjl 494, L199+

\bibitem[\protect\astroncite{{Engels} and {Bunzel}}{2015}]{Engels15}
{Engels} D., {Bunzel} F. (2015) \aap 582, A68

\bibitem[\protect\astroncite{{Fiege} and {Pudritz}}{2000}]{Fiege00}
{Fiege} J.D., {Pudritz} R.E. (2000) \mnras 311, 85

\bibitem[\protect\astroncite{{Fischera} and {Martin}}{2012}]{Fischera12}
{Fischera} J., {Martin} P.G. (2012) \aap 542, A77

\bibitem[\protect\astroncite{{Fujisawa} et~al.}{2015}]{Fujisawa15}
{Fujisawa} K., {Yonekura} Y., {Sugiyama} K., et~al. (2015) The Astronomer's
  Telegram 8286

\bibitem[\protect\astroncite{{Fukui} et~al.}{2015}]{Fukui15}
{Fukui} Y., {Harada} R., {Tokuda} K., et~al. (2015) \apj 807, L4

\bibitem[\protect\astroncite{{Fuller} and {Myers}}{1992}]{DCDCL7}
{Fuller} G.A., {Myers} P.C. (1992) \apj 384, 523

\bibitem[\protect\astroncite{{Gerner} et~al.}{2015}]{Gerner15}
{Gerner} T., {Shirley} Y.L., {Beuther} H., et~al. (2015) \aap 579, A80

\bibitem[\protect\astroncite{{Gibson} et~al.}{2009}]{Gibson09}
{Gibson} D., {Plume} R., {Bergin} E., et~al. (2009) \apj 705, 123

\bibitem[\protect\astroncite{{Goodman} et~al.}{1990}]{Goodman90}
{Goodman} A.A., {Bastien} P., {Myers} P.C., et~al. (1990) \apj 359, 363

\bibitem[\protect\astroncite{{Goodman} et~al.}{1993}]{DCDCL8}
{Goodman} A.A., {Benson} P.J., {Fuller} G.A., et~al. (1993) \apj 406, 528

\bibitem[\protect\astroncite{{Hacar} et~al.}{2017}]{Hacar17}
{Hacar} A., {Alves} J., {Tafalla} M., et~al. (2017) \aap 602, L2

\bibitem[\protect\astroncite{{Hacar} et~al.}{2018}]{Hacar18}
{Hacar} A., {Tafalla} M., {Forbrich} J., et~al. (2018) \aap 610, A77

\bibitem[\protect\astroncite{{Hacar} et~al.}{2013}]{Hacar13}
{Hacar} A., {Tafalla} M., {Kauffmann} J., et~al. (2013) \aap 554, A55

\bibitem[\protect\astroncite{{Hartmann}}{2002}]{Hartmann02}
{Hartmann} L. (2002) \apj 578, 914

\bibitem[\protect\astroncite{{Hunter} et~al.}{2017}]{Hunter17}
{Hunter} T.R., {Brogan} C.L., {MacLeod} G., et~al. (2017) \apj 837, L29

\bibitem[\protect\astroncite{{Hunter} et~al.}{2018}]{Hunter18}
{Hunter} T.R., {Brogan} C.L., {MacLeod} G.C., et~al. (2018) \apj 854, 170

\bibitem[\protect\astroncite{{Inoue} et~al.}{2018}]{Inoue18}
{Inoue} T., {Hennebelle} P., {Fukui} Y., et~al. (2018) Publications of the
  Astronomical Society of Japan 70, S53

\bibitem[\protect\astroncite{{Jackson} et~al.}{2006}]{Jackson06}
{Jackson} J.M., {Rathborne} J.M., {Shah} R.Y., et~al. (2006) \apjs 163, 145

\bibitem[\protect\astroncite{{Joncas} et~al.}{1992}]{Joncas92}
{Joncas} G., {Durand} D., {Roger} R.S. (1992) \apj 387, 591

\bibitem[\protect\astroncite{{Kainulainen} et~al.}{2016}]{Kainulainen16}
{Kainulainen} J., {Hacar} A., {Alves} J., et~al. (2016) \aap 586, A27

\bibitem[\protect\astroncite{{Kainulainen} et~al.}{2017}]{Kainulainen17}
{Kainulainen} J., {Stutz} A.M., {Stanke} T., et~al. (2017) \aap 600, A141

\bibitem[\protect\astroncite{{Kang} et~al.}{2017}]{Kang17}
{Kang} S.J., {Kerton} C.R., {Choi} M., et~al. (2017) \apj 845, 21

\bibitem[\protect\astroncite{{Kirk} et~al.}{2005}]{Kirk05}
{Kirk} J.M., {Ward-Thompson} D., {Andr{\'e}} P. (2005) \mnras 360, 1506

\bibitem[\protect\astroncite{{Kirsanova} et~al.}{2017}]{Kirsanova17}
{Kirsanova} M.S., {Salii} S.V., {Sobolev} A.M., et~al. (2017) Open Astronomy
  26, 99

\bibitem[\protect\astroncite{{Lada}}{1987}]{Lada87}
{Lada} C.J. (1987) In: Star Forming Regions, eds. M.~{Peimbert}, J.~{Jugaku},
  vol. 115 of IAU Symposium. 1

\bibitem[\protect\astroncite{{Ladd} et~al.}{1994}]{DCDCL10}
{Ladd} E.F., {Myers} P.C., {Goodman} A.A. (1994) \apj 433, 117

\bibitem[\protect\astroncite{{Lapinov} et~al.}{1998}]{Lapinov98}
{Lapinov} A.V., {Schilke} P., {Juvela} M., et~al. (1998) \aap 336, 1007

\bibitem[\protect\astroncite{{Larson}}{1981}]{Larson81}
{Larson} R.B. (1981) \mnras 194, 809

\bibitem[\protect\astroncite{{Larson}}{1985}]{Larson85}
{Larson} R.B. (1985) \mnras 214, 379

\bibitem[\protect\astroncite{{Li} et~al.}{2016}]{Li16}
{Li} G.X., {Urquhart} J.S., {Leurini} S., et~al. (2016) \aap 591, A5

\bibitem[\protect\astroncite{{Lintott} et~al.}{2005}]{Lintott05}
{Lintott} C.J., {Viti} S., {Rawlings} J.M.C., et~al. (2005) \apj 620, 795

\bibitem[\protect\astroncite{{Liu} et~al.}{2018}]{Liu18}
{Liu} S.Y., {Su} Y.N., {Zinchenko} I., et~al. (2018) \apj 863, L12

\bibitem[\protect\astroncite{{Liu} et~al.}{2020}]{Liu20}
{Liu} S.Y., {Su} Y.N., {Zinchenko} I., et~al. (2020) \apj 904, 181

\bibitem[\protect\astroncite{{Louvet}}{2018}]{Louvet18}
{Louvet} F. (2018) In: SF2A-2018: Proceedings of the Annual meeting of the
  French Society of Astronomy and Astrophysics. Di

\bibitem[\protect\astroncite{{Lu} et~al.}{2018}]{Lu18}
{Lu} X., {Zhang} Q., {Liu} H.B., et~al. (2018) \apj 855, 9

\bibitem[\protect\astroncite{{Malafeev} and {Zinchenko}}{2006}]{Malafeev06}
{Malafeev} S.Y., {Zinchenko} I.I. (2006) In: "Star Formation in the Galaxy and
  Beyond, Proceedings of the Conference "Star Formation in the Galaxy and
  Beyond", held in Moscow, Russia, 17-18 April 2006. Publisher: Moscow:
  Yanus-K, 2006. Edited by D. S. Wiebe and M. S. Kirsanova, ISBN: 05803703265
  (in Russian)". 29

\bibitem[\protect\astroncite{{Malafeev} et~al.}{2005}]{Malafeev05}
{Malafeev} S.Y., {Zinchenko} I.I., {Pirogov} L.E., et~al. (2005) Astronomy
  Letters 31, 239

\bibitem[\protect\astroncite{{Mallick} et~al.}{2013}]{Mallick13}
{Mallick} K.K., {Kumar} M.S.N., {Ojha} D.K., et~al. (2013) \apj 779, 113

\bibitem[\protect\astroncite{{McKee} and {Ostriker}}{2007}]{McKee07}
{McKee} C.F., {Ostriker} E.C. (2007) \araa 45, 565

\bibitem[\protect\astroncite{{Men'shchikov}}{2013}]{Menshchikov13}
{Men'shchikov} A. (2013) \aap 560, A63

\bibitem[\protect\astroncite{{Meyer} et~al.}{2017}]{Meyer17}
{Meyer} D.M.A., {Vorobyov} E.I., {Kuiper} R., et~al. (2017) \mnras 464, L90

\bibitem[\protect\astroncite{{Molinari} et~al.}{1996}]{Molinari96}
{Molinari} S., {Brand} J., {Cesaroni} R., et~al. (1996) \aap 308, 573

\bibitem[\protect\astroncite{{Molinari} et~al.}{2016}]{Molinari16}
{Molinari} S., {Schisano} E., {Elia} D., et~al. (2016) \aap 591, A149

\bibitem[\protect\astroncite{{Moscadelli} et~al.}{2017}]{Moscadelli17}
{Moscadelli} L., {Sanna} A., {Goddi} C., et~al. (2017) \aap 600, L8

\bibitem[\protect\astroncite{{Motte} et~al.}{2018}]{Motte18}
{Motte} F., {Bontemps} S., {Louvet} F. (2018) \araa 56, 41

\bibitem[\protect\astroncite{{Myers}}{1983}]{DCDCL3}
{Myers} P.C. (1983) \apj 270, 105

\bibitem[\protect\astroncite{{Myers}}{1985}]{Myers85}
{Myers} P.C. (1985) In: Protostars and Planets II, eds. D.C. {Black}, M.S.
  {Matthews}. 81--103

\bibitem[\protect\astroncite{{Myers}}{2009}]{Myers09}
{Myers} P.C. (2009) \apj 700, 1609

\bibitem[\protect\astroncite{{Myers} and {Benson}}{1983}]{DCDCL2}
{Myers} P.C., {Benson} P.J. (1983) \apj 266, 309

\bibitem[\protect\astroncite{{Myers} et~al.}{1991}]{DCDCL6}
{Myers} P.C., {Fuller} G.A., {Goodman} A.A., et~al. (1991) \apj 376, 561

\bibitem[\protect\astroncite{{Myers} et~al.}{1988}]{DCDCL5}
{Myers} P.C., {Heyer} M., {Snell} R.L., et~al. (1988) \apj 324, 907

\bibitem[\protect\astroncite{{Myers} et~al.}{1983}]{DCDCL1}
{Myers} P.C., {Linke} R.A., {Benson} P.J. (1983) \apj 264, 517

\bibitem[\protect\astroncite{{Nakamura} et~al.}{2014}]{Nakamura14}
{Nakamura} F., {Sugitani} K., {Tanaka} T., et~al. (2014) \apj 791, L23

\bibitem[\protect\astroncite{{Offner} et~al.}{2014}]{Offner14}
{Offner} S.S.R., {Clark} P.C., {Hennebelle} P., et~al. (2014) In: Protostars
  and Planets VI, eds. H.~{Beuther}, R.S. {Klessen}, C.P. {Dullemond},
  T.~{Henning}. 53

\bibitem[\protect\astroncite{{Ojha} et~al.}{2011}]{Ojha11}
{Ojha} D.K., {Samal} M.R., {Pandey} A.K., et~al. (2011) \apj 738, 156

\bibitem[\protect\astroncite{{Ostriker}}{1964}]{Ostriker64}
{Ostriker} J. (1964) \apj 140, 1056

\bibitem[\protect\astroncite{{Perault} et~al.}{1996}]{Perault96}
{Perault} M., {Omont} A., {Simon} G., et~al. (1996) \aap 315, L165

\bibitem[\protect\astroncite{{Peretto} et~al.}{2014}]{Peretto14}
{Peretto} N., {Fuller} G.A., {Andr{\'e}} P., et~al. (2014) \aap 561, A83

\bibitem[\protect\astroncite{{Pilbratt} et~al.}{2010}]{Pilbratt10}
{Pilbratt} G.L., {Riedinger} J.R., {Passvogel} T., et~al. (2010) \aap 518, L1

\bibitem[\protect\astroncite{{Pillai} et~al.}{2006}]{Pillai06}
{Pillai} T., {Wyrowski} F., {Carey} S.J., et~al. (2006) \aap 450, 569

\bibitem[\protect\astroncite{{Pirogov} et~al.}{1996}]{Pirogov96}
{Pirogov} L., {Lapinov} A., {Zinchenko} I., et~al. (1996) Astronomical and
  Astrophysical Transactions 11, 287

\bibitem[\protect\astroncite{{Pirogov} et~al.}{2013}]{Pirogov13}
{Pirogov} L., {Ojha} D.K., {Thomasson} M., et~al. (2013) \mnras 436, 3186

\bibitem[\protect\astroncite{{Pirogov} et~al.}{2003}]{Pirogov03}
{Pirogov} L., {Zinchenko} I., {Caselli} P., et~al. (2003) \aap 405, 639

\bibitem[\protect\astroncite{{Pirogov} et~al.}{2007}]{Pirogov07}
{Pirogov} L., {Zinchenko} I., {Caselli} P., et~al. (2007) \aap 461, 523

\bibitem[\protect\astroncite{{Pirogov}}{2009}]{Pirogov09}
{Pirogov} L.E. (2009) Astronomy Reports 53, 1127

\bibitem[\protect\astroncite{{Pirogov}}{2022}]{Pirogov22x}
{Pirogov} L.E. (2022) arXiv e-prints , arXiv:2208.03166

\bibitem[\protect\astroncite{{Pirogov} and {Zinchenko}}{1998}]{Pirogov98}
{Pirogov} L.E., {Zinchenko} I.I. (1998) Astronomy Reports 42, 11

\bibitem[\protect\astroncite{{Plume} et~al.}{1997}]{Plume97}
{Plume} R., {Jaffe} D.T., {Evans} II N.J., et~al. (1997) \apj 476, 730

\bibitem[\protect\astroncite{{Ragan} et~al.}{2009}]{Ragan09}
{Ragan} S.E., {Bergin} E.A., {Gutermuth} R.A. (2009) \apj 698, 324

\bibitem[\protect\astroncite{{Rathborne} et~al.}{2006}]{Rathborne06}
{Rathborne} J.M., {Jackson} J.M., {Simon} R. (2006) \apj 641, 389

\bibitem[\protect\astroncite{{Richards} et~al.}{2012}]{Richards12}
{Richards} E.E., {Lang} C.C., {Trombley} C., et~al. (2012) \aj 144, 89

\bibitem[\protect\astroncite{{Rosen} et~al.}{2020}]{Rosen20}
{Rosen} A.L., {Offner} S.S.R., {Sadavoy} S.I., et~al. (2020) \ssr 216, 62

\bibitem[\protect\astroncite{{Roueff} et~al.}{2005}]{Roueff05}
{Roueff} E., {Lis} D.C., {van der Tak} F.F.S., et~al. (2005) \aap 438, 585

\bibitem[\protect\astroncite{{Roueff} et~al.}{2007}]{Roueff07}
{Roueff} E., {Parise} B., {Herbst} E. (2007) \aap 464, 245

\bibitem[\protect\astroncite{{Russeil} et~al.}{2007}]{Russeil07}
{Russeil} D., {Adami} C., {Georgelin} Y.M. (2007) \aap 470, 161

\bibitem[\protect\astroncite{{Ryabukhina} and {Zinchenko}}{2021}]{Ryabukhina21}
{Ryabukhina} O.L., {Zinchenko} I.I. (2021) \mnras 505, 726

\bibitem[\protect\astroncite{{Ryabukhina} et~al.}{2018}]{Ryabukhina18}
{Ryabukhina} O.L., {Zinchenko} I.I., {Samal} M.R., et~al. (2018) Research in
  Astronomy and Astrophysics 18, 095

\bibitem[\protect\astroncite{{Sakai} et~al.}{2008}]{Sakai08}
{Sakai} T., {Sakai} N., {Kamegai} K., et~al. (2008) \apj 678, 1049

\bibitem[\protect\astroncite{{Sakamoto} et~al.}{1997}]{Sakamoto97}
{Sakamoto} S., {Hasegawa} T., {Handa} T., et~al. (1997) \apj 486, 276

\bibitem[\protect\astroncite{{Samal} et~al.}{2015}]{Samal15}
{Samal} M.R., {Ojha} D.K., {Jose} J., et~al. (2015) \aap 581, A5

\bibitem[\protect\astroncite{{S{\'a}nchez-Monge}
  et~al.}{2014}]{Sanchez-Monge14}
{S{\'a}nchez-Monge} {\'A}., {Beltr{\'a}n} M.T., {Cesaroni} R., et~al. (2014)
  \aap 569, A11

\bibitem[\protect\astroncite{{Schisano} et~al.}{2020}]{Schisano20}
{Schisano} E., {Molinari} S., {Elia} D., et~al. (2020) \mnras 492, 5420

\bibitem[\protect\astroncite{{Schuller} et~al.}{2009}]{Schuller09}
{Schuller} F., {Menten} K.M., {Contreras} Y., et~al. (2009) \aap 504, 415

\bibitem[\protect\astroncite{{Sridharan} et~al.}{2002}]{Sridharan02}
{Sridharan} T.K., {Beuther} H., {Schilke} P., et~al. (2002) \apj 566, 931

\bibitem[\protect\astroncite{{Stod{\'o}lkiewicz}}{1963}]{Stodolkiewicz63}
{Stod{\'o}lkiewicz} J.S. (1963) Acta Astronomica 13, 30

\bibitem[\protect\astroncite{{Tan} et~al.}{2014}]{Tan14}
{Tan} J.C., {Beltr{\'a}n} M.T., {Caselli} P., et~al. (2014) In: Protostars and
  Planets VI, eds. H.~{Beuther}, R.S. {Klessen}, C.P. {Dullemond},
  T.~{Henning}. 149

\bibitem[\protect\astroncite{{Torrelles} et~al.}{2011}]{Torrelles11}
{Torrelles} J.M., {Patel} N.A., {Curiel} S., et~al. (2011) \mnras 410, 627

\bibitem[\protect\astroncite{{Tritsis} and {Tassis}}{2016}]{Tritsis16}
{Tritsis} A., {Tassis} K. (2016) \mnras 462, 3602

\bibitem[\protect\astroncite{{Trofimova} et~al.}{2020}]{Trofimova20}
{Trofimova} E.A., {Zinchenko} I.I., {Zemlyanukha} P.M., et~al. (2020) Astronomy
  Reports 64, 244

\bibitem[\protect\astroncite{{Valdettaro} et~al.}{2001}]{Valdettaro01}
{Valdettaro} R., {Palla} F., {Brand} J., et~al. (2001) \aap 368, 845

\bibitem[\protect\astroncite{{Vasyunina} et~al.}{2009}]{Vasyunina09}
{Vasyunina} T., {Linz} H., {Henning} T., et~al. (2009) \aap 499, 149

\bibitem[\protect\astroncite{{Vasyunina} et~al.}{2011}]{Vasyunina11}
{Vasyunina} T., {Linz} H., {Henning} T., et~al. (2011) \aap 527, A88

\bibitem[\protect\astroncite{{Vilas-Boas} et~al.}{1994}]{DCDCL9}
{Vilas-Boas} J.W.S., {Myers} P.C., {Fuller} G.A. (1994) \apj 433, 96

\bibitem[\protect\astroncite{{Vilas-Boas} et~al.}{2000}]{DCDCL12}
{Vilas-Boas} J.W.S., {Myers} P.C., {Fuller} G.A. (2000) \apj 532, 1038

\bibitem[\protect\astroncite{{Walsh} et~al.}{1997}]{Walsh97}
{Walsh} A.J., {Hyland} A.R., {Robinson} G., et~al. (1997) \mnras 291, 261

\bibitem[\protect\astroncite{{Wang} et~al.}{2011}]{Wang11}
{Wang} Y., {Beuther} H., {Bik} A., et~al. (2011) \aap 527, A32

\bibitem[\protect\astroncite{{Ward-Thompson} et~al.}{2010}]{Ward-Thompson10}
{Ward-Thompson} D., {Kirk} J.M., {Andr{\'e}} P., et~al. (2010) \aap 518, L92

\bibitem[\protect\astroncite{{Wood} and {Churchwell}}{1989}]{Wood89}
{Wood} D.O.S., {Churchwell} E. (1989) \apj 340, 265

\bibitem[\protect\astroncite{{Wright} et~al.}{2010}]{Wright10}
{Wright} E.L., {Eisenhardt} P.R.M., {Mainzer} A.K., et~al. (2010) \aj 140,
  1868-1881

\bibitem[\protect\astroncite{{Zemlyanukha} et~al.}{2022}]{Zemlyanukha22}
{Zemlyanukha} P., {Zinchenko} I.I., {Dombek} E., et~al. (2022) \mnras 515, 2445

\bibitem[\protect\astroncite{{Zemlyanukha} et~al.}{2018}]{Zemlyanukha18}
{Zemlyanukha} P.M., {Zinchenko} I.I., {Salii} S.V., et~al. (2018) Astronomy
  Reports 62, 326

\bibitem[\protect\astroncite{{Zinchenko}}{1995}]{Zin95-2}
{Zinchenko} I. (1995) \aap 303, 554

\bibitem[\protect\astroncite{{Zinchenko} et~al.}{2009}]{Zin09}
{Zinchenko} I., {Caselli} P., {Pirogov} L. (2009) \mnras 395, 2234

\bibitem[\protect\astroncite{{Zinchenko} et~al.}{1994}]{Zin94}
{Zinchenko} I., {Forsstroem} V., {Lapinov} A., et~al. (1994) \aap 288, 601

\bibitem[\protect\astroncite{{Zinchenko} and {Henkel}}{2018}]{Zin-SO}
{Zinchenko} I., {Henkel} C. (2018) In: IAU Symposium, eds. M.~{Cunningham},
  T.~{Millar}, Y.~{Aikawa}, vol. 332 of IAU Symposium. 274--277

\bibitem[\protect\astroncite{{Zinchenko} et~al.}{2000}]{Zin00}
{Zinchenko} I., {Henkel} C., {Mao} R.Q. (2000) \aap 361, 1079

\bibitem[\protect\astroncite{{Zinchenko} et~al.}{1997}]{Zin97}
{Zinchenko} I., {Henning} T., {Schreyer} K. (1997) \aaps 124, 385

\bibitem[\protect\astroncite{{Zinchenko} et~al.}{2011}]{Zin11-iau}
{Zinchenko} I., {Kurtz} S., {Liu} S.Y., et~al. (2011) In: The Molecular
  Universe, eds. J.~{Cernicharo}, R.~{Bachiller}, vol. 280. 392

\bibitem[\protect\astroncite{{Zinchenko} et~al.}{2012}]{Zin12}
{Zinchenko} I., {Liu} S.Y., {Su} Y.N., et~al. (2012) \apj 755, 177

\bibitem[\protect\astroncite{{Zinchenko} et~al.}{2015}]{Zin15}
{Zinchenko} I., {Liu} S.Y., {Su} Y.N., et~al. (2015) \apj 810, 10

\bibitem[\protect\astroncite{{Zinchenko} et~al.}{2017{\natexlab{a}}}]{Zin17}
{Zinchenko} I., {Liu} S.Y., {Su} Y.N., et~al. (2017{\natexlab{a}}) \aap 606, L6

\bibitem[\protect\astroncite{{Zinchenko}
  et~al.}{2018{\natexlab{a}}}]{Zin18-raa}
{Zinchenko} I., {Liu} S.Y., {Su} Y.N., et~al. (2018{\natexlab{a}}) Research in
  Astronomy and Astrophysics 18, 093

\bibitem[\protect\astroncite{{Zinchenko}
  et~al.}{2018{\natexlab{b}}}]{Zin18-iau}
{Zinchenko} I., {Liu} S.Y., {Su} Y.N., et~al. (2018{\natexlab{b}}) In: IAU
  Symposium, eds. M.~{Cunningham}, T.~{Millar}, Y.~{Aikawa}, vol. 332. 270--273

\bibitem[\protect\astroncite{{Zinchenko} et~al.}{1995}]{Zin95}
{Zinchenko} I., {Mattila} K., {Toriseva} M. (1995) \aaps 111, 95

\bibitem[\protect\astroncite{{Zinchenko}
  et~al.}{2017{\natexlab{b}}}]{Zin17-sao}
{Zinchenko} I., {Ojha} D.K., {Pirogov} L., et~al. (2017{\natexlab{b}}) In:
  Stars: From Collapse to Collapse, eds. Y.Y. {Balega}, D.O. {Kudryavtsev},
  I.I. {Romanyuk}, I.A. {Yakunin}, vol. 510 of Astronomical Society of the
  Pacific Conference Series. 9

\bibitem[\protect\astroncite{{Zinchenko} et~al.}{2005}]{Zin05}
{Zinchenko} I., {Pirogov} L., {Caselli} P., et~al. (2005) In: Massive Star
  Birth: A Crossroads of Astrophysics, eds. R.~{Cesaroni}, M.~{Felli},
  E.~{Churchwell}, M.~{Walmsley}, vol. 227. 92--97

\bibitem[\protect\astroncite{{Zinchenko} et~al.}{1998}]{Zin98}
{Zinchenko} I., {Pirogov} L., {Toriseva} M. (1998) \aaps 133, 337

\bibitem[\protect\astroncite{{Zinchenko} et~al.}{2021}]{Zinchenko21}
{Zinchenko} I.I., {Dewangan} L.K., {Baug} T., et~al. (2021) \mnras 506, L45

\bibitem[\protect\astroncite{{Zinchenko} et~al.}{1990}]{Zin90}
{Zinchenko} I.I., {Krasil'Nikov} A.A., {Kukina} E.P., et~al. (1990) \azh 67,
  908

\bibitem[\protect\astroncite{{Zinchenko} et~al.}{1989}]{Zin89}
{Zinchenko} I.I., {Lapinov} A.V., {Pirogov} L.E. (1989) \azh 66, 1142

\bibitem[\protect\astroncite{{Zinchenko} et~al.}{2020}]{Zin20}
{Zinchenko} I.I., {Liu} S.Y., {Su} Y.N., et~al. (2020) \apj 889, 43

\bibitem[\protect\astroncite{{Zinchenko} et~al.}{2022}]{Zinchenko22}
{Zinchenko} I.I., {Pazukhin} A.G., {Trofimova} E.A., et~al. (2022) PoS
  MUTO2022, 038

\bibitem[\protect\astroncite{Zinnecker and Yorke}{2007}]{Zinnecker07}
Zinnecker H., Yorke H.W. (2007) \araa 45, 481

\end{thebibliography}
\end{document}